\documentclass[a4paper,11pt]{article}
\pdfoutput=1
\usepackage{jheppub}

\usepackage[T1]{fontenc}
\usepackage{amsfonts}
\usepackage{amsmath}
\usepackage{amssymb}
\usepackage{multirow}
\usepackage{graphicx}
\usepackage{subfig}

\usepackage{array,ragged2e}

\usepackage{scrextend}

\usepackage{tikz-cd}

\usepackage{mathtools}

\usepackage{xfrac}

\DeclareMathOperator*{\RE}{Re}
\DeclareMathOperator*{\IM}{Im}

\DeclareMathOperator*{\Jac}{Jac}

\DeclareMathOperator{\Th}{Th}

\newcommand{\F}{{\mathbb{F}}}
\newcommand{\Z}{{\mathbb{Z}}}
\newcommand{\C}{{\mathbb{C}}}

\newcommand{\thetaUD}[2]{{\renewcommand{\arraystretch}{1}\theta\begin{bmatrix}#1 \\ #2 \end{bmatrix}}}

\newcommand{\vek}[1]{{\boldsymbol{#1}}}

\numberwithin{equation}{section}

\usepackage{graphicx}

\begin{document}

\title{Classical Codes and Chiral CFTs at Higher Genus}
\author[a]{Johan Henriksson,}
\emailAdd{johan.henriksson@df.unipi.it}
\affiliation[a]{Department of Physics, University of Pisa and INFN, \\Largo Pontecorvo 3, I-56127 Pisa, Italy}
\author[b]{Ashish Kakkar,}
\emailAdd{ashish.kakkar@uky.edu}
\affiliation[b]{Department of Physics and Astronomy, University of Kentucky,\\ Lexington, KY 40506, U.S.A}
\author[a]{Brian McPeak}
\emailAdd{brian.mcpeak@df.unipi.it}

\abstract{Higher genus modular invariance of two-dimensional conformal field theories (CFTs) is a largely unexplored area. In this paper, we derive explicit expressions for the higher genus partition functions of a specific class of CFTs: code CFTs, which are constructed using classical error-correcting codes. In this setting, the $\mathrm{Sp}(2g,\mathbb Z)$ modular transformations of genus $g$ Riemann surfaces can be recast as a simple set of linear maps acting on $2^g$ polynomial variables, which comprise an object called the code enumerator polynomial. The CFT partition function is directly related to the enumerator polynomial, meaning that solutions of the linear constraints from modular invariance immediately give a set of seemingly consistent partition functions at a given genus. We then find that higher genus constraints, plus consistency under degeneration limits of the Riemann surface, greatly reduces the number of possible code CFTs. This work provides a step towards a full understanding of the constraints from higher genus modular invariance on 2d CFTs.}

\maketitle
\flushbottom

\section{Introduction}

Two-dimensional conformal field theories are among the best understood quantum field theories, and yet a lot remains unknown. This is especially true for theories with central charge $c> 1$, where unitarity allows for infinite-dimensional representations of the Virasoro algebra. The resulting theories are, in general, much more complicated than those theories with $c<1$. One powerful method for studying 2d conformal field theories (CFTs) is the modular bootstrap \cite{Hellerman:2009bu}, which constrains the torus partition function based on the requirement that it be invariant under the $\mathrm{PSL}(2,\mathbb{Z})$ group of modular transformations. The constraints from genus 1 modular invariance have been used to derive universal bounds on the spectrum, including bounds on the dimension gap, twist gap, and operator degeneracies \cite{Cardy:1986ie,Hellerman:2010qd, Keller:2012mr, Friedan:2013cba, Qualls:2013eha, Kim:2015oca, Collier:2016cls, Benjamin:2016fhe, Dyer:2017rul, Anous:2018hjh, Bae:2018qym, Lin:2019kpn, Afkhami-Jeddi:2019zci, Hartman:2019pcd}.

The torus partition function fully specifies the spectrum of the theory, but contains no information about the dynamics, \emph{i.e.}\ the OPE coefficients. For this reason, it is unable to completely define a theory. This is clear from specific cases where different theories have the same genus 1 partition function, the most famous of which arise from Milnor's example of two isospectral self-dual even lattices in 16 dimensions which define isospectral CFTs with $c = 16$ \cite{Milnor}. Perhaps more surprisingly, there exist examples of modular invariant functions, decomposable in Virasoro characters with non-negative integer multiplicities, which do not correspond to any known CFT; for an example in the non-chiral case, see section~6 of \cite{Dymarsky:2020qom}. The dynamical information captured by the OPE coefficients can, in principle, be addressed using the full conformal bootstrap method \cite{Rattazzi:2008pe} to bound sphere four-point functions using crossing symmetry, however these constraints are not easily combined with those of modular invariance. 

Considering higher genus partition functions might offer a way to bound the spectrum and OPE coefficients at the same time. It was argued in \cite{Moore:1988qv} that consistency of 2d CFTs with crossing symmetry and modular invariance implies that the theory can be consistently defined on Riemann surfaces of any genus. The same Riemann surface can be obtained by sewing together simple three-holed spheres in different ways -- higher genus consistency essentially means that the resulting partition function must not depend on the sewing procedure. This leads to the requirement that the partition function is invariant under the full genus-$g$ modular group $\mathrm{Sp}(2g, \mathbb{Z})$. Unlike the torus partition function, the higher genus partition functions do contain information about the OPE coefficients. This leads to the natural question: to what degree does the set of higher genus partition functions characterize a 2d CFT? The view that a 2d CFT may be defined by its vacuum amplitudes for all genera, originally advanced in \cite{Friedan:1986ua}, has been addressed more recently in \cite{Gaberdiel:2009rd, Gaberdiel:2010jf}. Other recent interest in deriving universal bounds from higher genus modular invariance include \cite{Keller:2017iql, Cho:2017fzo, Cardy:2017qhl}. 

In this paper, we will use 2d CFTs defined by error-correcting codes (ECCs) to study higher genus modular invariance. Simply speaking, an $n$-dimensional code $\mathcal C$ is a collection of binary vectors, ``codewords,'' of length $n$. These vectors can be visualized as vertices on a unit hypercube. Importantly, this means that an error correcting code can be used to define a lattice, simply by identifying the vectors in the code with the unit cell of the lattice (this is the so-called ``Construction A,'' of Leech and Sloane \cite{Leech1993}). This, in turn, allows us to define a chiral CFT with central charge $c=n$ as the compactification of $n$ free bosons on this lattice \cite{Dolan1996}. The relationship between code and CFTs has been known for a long time \cite{Dolan:1989kf,Dolan:1994st}. Recently it has been shown \cite{Dymarsky:2020bps, Dymarsky:2020qom} that the correspondence between classical codes and chiral CFTs has a generalization to quantum error correcting codes and full non-chiral CFTs.

A key element of the relationship between codes and CFTs is that the CFT partition function is given by the code enumerator polynomial $W_{\mathcal C}$, which counts the degeneracy of codewords much like the partition function counts the states of the theory. The relation between the enumerator and the torus partition function is\footnote{The denominator $\Phi(\tau)$ is not theory-dependent, so it will not enter into our discussion. In what follows, we will sometimes use ``partition function'' to refer to the numerator only.} 
\begin{align}
    Z(\tau) = \frac{W_{\mathcal C}(\vartheta_0, \vartheta_1)}{\Phi(\tau)} \,  ,
    \label{eq:basic}
\end{align}
where $\vartheta_i=\vartheta_i(\tau)$ are Jacobi theta functions. In this paper, we show that the relationship~\eqref{eq:basic} extends to higher genus partition functions, which are computed by the ``higher-weight'' enumerator polynomials. The higher-weight enumerator polynomials have been known in the literature on ECCs since \cite{MacWilliams1972}, but as far as we know, the connection to higher genus partition functions has never been pointed out.

Code CFTs provide an interesting setting for studying higher genus modular invariance for two main reasons. The first is that a number of codes are known explicitly, allowing us to construct higher genus partition functions of true CFTs directly by computing their higher-weight enumerator polynomial. The second is that the higher genus modular transformations act \emph{linearly} on partition functions which take the form of code enumerator polynomials.\footnote{In fact, any lattice CFT has partition functions which can be written in the form of a code enumerator polynomial, and thus must transform the same way.} This means that we can actually solve the transformations explicitly. By requiring that the enumerator polynomial has positive integer coefficients, we can list every partition function which could possibly derive from an error-correcting code. Finally, we impose a further constraint, called ``factorization limits:'' when the higher genus Riemann surface degenerates into two lower-genus Riemann surfaces connected by an infinitely thin, long tube, then the higher genus partition function must factorize into the products of lower-genus partition functions. 

For cases where the full set of codes is known, such as for $c = 24$, we find that this procedure correctly reproduces every code partition function, plus a number of ``fake'' partition functions, which are known not to correspond to codes. Whether or not they correspond to CFTs at all is an open question. Interestingly, imposing constraints from higher genus decreases the number of fake partition functions. For example, at $c = 24$, there are only 9 codes. At genus 1, there are 190 partition functions consistent with modular invariance. However, we find that only 29 of these partition functions can arise from modular invariant genus 2 partition functions via factorization. Only 21 can arise from genus 3 partition functions. We speculate that performing the algorithm laid out in this paper for arbitrarily high genus would eliminate every partition function except those arising from ECCs. 

For genus $g \leqslant 3$ and central charge $c = 24$, we can look at the location of the true theories inside the space of all allowed theories, revealing an interesting geometric structure. The requirement that the partition functions have positive integer coefficients means that the allowed partition functions form a discrete set indexed by a few coefficients which obey simple linear inequalities. More specifically, these coefficients live in a polytope whose dimension is equal to one less than the number of independent Siegel modular forms with weight $c/2$. This dimension grows as the genus grows, but the higher genus polytope can always be projected to the lower-genus space, and its projection must be inside the polytope defined by the lower-genus constraints. This means that each genus gives stricter constraints than the previous one, at least for $g \leqslant 3$.

The outline of the paper is as follows. In section \ref{sec:review}, we review the correspondence described above in the case of chiral CFTs at genus 1. In section \ref{sec:Higher Genus}, we show how to extend the correspondence between codes and CFTs to higher genus, focusing on the higher-weight enumerator polynomial and how it is constrained by the symmetries and factorization properties required by the CFT. We also comment on some issues related to genus $g>3$. In section \ref{sec:enumeration}, we show explicitly how to use these constraints to determine every allowed enumerator polynomial for a given genus $g$ and dimension $c$. We do this for $c$ up to $48$ for genus $2$ and $c$ up to $24$ for genus 3. We show where the real ECCs lie in the space of allowed enumerator polynomials. We also use our results to fix the genus 2 Siegel modular forms in terms of the polynomial variables of ECCs.

\section{Review: Classical Codes and Chiral CFTs}
\label{sec:review}

In this section, we will review the relation between classical error-correcting codes and two-dimensional chiral CFTs. See table~\ref{tab:introtable} for a summary. 

\begin{table}
\centering
\caption{Dictionary relating error correcting codes and CFTs.}\label{tab:introtable}
{\renewcommand{\arraystretch}{1.75}
\begin{tabular}{>{\RaggedRight\arraybackslash}p{4.5cm}>{\RaggedRight\arraybackslash}p{4.5cm}>{\RaggedRight\arraybackslash}p{4.5cm}}
Code $\mathcal{C}$ & Lattice $\Lambda$ & Chiral CFT
\\\hline \hline
code dimension $n$ & dimension $n$ of $\mathbb{R}^n$ & central charge $c=n$
\\\hline
codewords $c\in \F_2^n$ & lattice vectors $\vec{v}$ & states 
\\\hline
weight $g$ enumerator polynomial $W_\mathcal{C}^{(g)}$ & genus g theta series  $\Theta_{\Lambda}^{g}$ & genus $g$ partition function $Z^{(g)}$
\\\hline
code variables $x_i=x_0,x_1, \ldots $ & theta constants $\vartheta_i(\Omega)$ \eqref{eq:thetaconstantsdefn} & theta constants $\vartheta_i(\Omega)$ 
\\\hline
Hamming weight $w(c)$ &  vector length $\lVert v \rVert$ & dimension $h$
\\ \hline
doubly-even self dual & even self-dual & modular invariant
\\\hline
\end{tabular}
}
\end{table}

\subsection{Error-correcting codes and enumerator polynomials}

In this paper, we are considering the problem of listing all possible enumerator polynomials by generating all polynomials consistent with a set of particular symmetries. So let us first review how classical error-correcting codes define enumerator polynomials, and their relation to CFTs. For a more extensive recent review of these topics, see \cite{Dymarsky:2020qom}. For an earlier review, see \cite{Elkies_lattices_i,Elkies_lattices_ii}.

A \emph{binary code} $\mathcal{C}$ is a set of codewords $c\in \F_2^n$, which are length $n$ vectors over $\F_2$, the finite field of two elements. The idea of an error-correcting code is to facilitate communication over a noisy channel: if a message is encoded, the message may still be readable even when a certain number of flipped bits (errors) occur between the sender and receiver. The simplest example is the repetition code. The length $3$ repetition code is defined by the encoding
\begin{align}
\begin{split}
    (0)  \, & \to \, (000) \, , \\
    (1)  \, & \to \, (111) \, .
    \label{eq:3repcode}
\end{split}
\end{align}
Suppose that each bit has a $10\%$ chance of being flipped when transmitting the message. If the original message is encoded as $(1)$, then it has a $90\%$ chance of being read correctly by the receiver. But if the sender instead transmits $(111)$, then the receiver may interpret the original message $(1)$ from any of $(111)$ (no errors) or $(110)$, $(101)$, or $(011)$ (one error). If there are two or three errors, however, the receiver will incorrectly reconstruct the original message. So the receiver has a $97.2\% $ chance of correctly interpreting the message.  

In the example above, the receiver reconstructed the message by replacing a three-bit signal, (\emph{e.g.}\ $(110)$), with the ``nearest'' codeword (\emph{e.g.}\ $(111)$). Nearest, in this case, means with the lowest Hamming distance $d$. For two vectors $c_1, \, c_2 \in \F_2^n$, the Hamming distance is defined as the number of bits in $c_1$ and $c_2$ which are different, equivalently $|c_1- c_2|$. 
The Hamming distance for a code $\mathcal{C}$ is defined as the minimum of the Hamming distance between each distinct elements $c\in\mathcal{C}$. The Hamming distance gives a measure of error-correcting abilities; a code with Hamming distance $d$ can correct errors in up to $t = \left \lfloor{(d - 1)/2}\right \rfloor $ errors. A code which has $2^k$ codewords living in $\F_2^n$, and which has Hamming distance $d$, is called a $[n, k, d]$ code. For example, the repetition code defined in \eqref{eq:3repcode} is a $[3, 1, 3]$ code, and is thus able to correct up to $1$ error. 

The key to error-correcting codes is redundancy -- the more redundancy there is, the more errors are allowed before the receiver will misinterpret the message. For example, if we used $(00000)$ and $(11111)$ as the codewords, we would find $99.144 \%$ chance of guessing the correct message. This would correspond to a $[5,1,5]$ code, and could therefore detect up to $2$ errors. A central question in coding theory is how to design codes which give the highest error-correcting ability, represented by $d$, for fixed values of $n$ and $k$. 

\subsubsection{Doubly-even self-dual codes}

To proceed, we will need to invoke a few more definitions from coding theory. A code is \emph{linear} if the sum of two codewords is a codeword. From here on, we will only consider linear codes. For a linear code $\mathcal{C}$, we can define the \emph{dual} code $\mathcal{C}^\perp$
\begin{align}
    \mathcal{C}^\perp = \left\{ \tilde c \ | \ \tilde c \cdot c \equiv 0 \, (\text{mod } 2)\,  \forall \, c \in \mathcal{C} \right\}.
\end{align}
If $\mathcal{C}$ is a type $[n, k, d]$ code, then its dual $\mathcal{C}^\perp$ will be of type $[n, n - k, d']$. A code for which $\mathcal{C} = \mathcal{C}^\perp$ is called \emph{self-dual}. Clearly, this is only possible when $n = 2 k$.

We also need to define the \emph{Hamming weight}: for a vector $c \in \F_2^n$, the Hamming weight $w(c)$ is simply the number of $1$s in $c$: $w(c)=|c|$. A linear code is called \emph{even} if the Hamming weight of all codewords is divisible by two. It is \emph{doubly-even} if all Hamming weights are divisible by four. Looking ahead, doubly-even self-dual codes are particularly relevant since they define lattices which are even and self-dual. Such lattices define chiral CFTs according to a construction by Dolan, Goddard and Montague \cite{Dolan:1989kf,Dolan:1994st}.

\subsubsection{Enumerator polynomials}

A convenient quantity for describing codes is the (weight one) \emph{enumerator polynomial}, defined by
\begin{align}
    W_\mathcal{C}(x_0, x_1) = \sum_{c \in \mathcal C} x_0^{n - w(c)} x_1^{w(c)} \, .
\end{align}
The enumerator polynomial essentially counts all of the codewords with given Hamming weight. Linear codes always contain the codeword $\vec 0$, so the enumerator polynomial always contains the monomial $x_0^n$ with coefficient 1. The coefficients of each monomial are all positive integers because they are degeneracies, and they must add up to $2^k$. For example, the simplest self-dual double even code, the Hamming $[8, 4, 4]$ code has the enumerator polynomial
\begin{align}
    W_{\text{Hamming}}(x_0, x_1) = x_0^8 + 14 x_0^4 x_1^4 + x_1^8 \, .
\end{align}
The enumerator polynomial describes a code in terms the Hamming weights of its elements much like the partition function describes a CFT in terms of the dimensions of its states; later, we will describe how this analogy is precise for lattice CFTs. This analogy also serves to emphasize that it is not, in general, possible to extract the entire code its (genus 1) enumerator polynomial. 

Code duality $\mathcal{C} \mapsto \mathcal{C}^\perp$ can be expressed as a transformation on the enumerator polynomials according to the MacWilliams identity \cite{Macwilliams1963}:
\begin{align}
    x_0 \ \mapsto\  \frac{x_0 + x_1}{\sqrt 2} \, , \qquad x_1 \ \mapsto \ \frac{x_0 - x_1}{\sqrt 2} \, .
    \label{eq:EP_s}
\end{align}
Enumerator polynomials for self-dual codes must therefore be invariant under this transformation. Furthermore, doubly-even codes must be invariant under 
\begin{align}
    x_1\ \mapsto\  i x_1 \, .
    \label{eq:EP_t}
\end{align}
The transformations \eqref{eq:EP_s} and \eqref{eq:EP_t} are directly related to the $S$ and $T$ modular transformations on CFT partition functions. They represent a powerful set of constraints on the possible enumerator polynomials. This idea is central to the method of this paper. At a given polynomial degree, we can solve these constraints to find the most general set of polynomials which are invariant under these transformations. If we require that the coefficients are positive and integer, there are finite number of solutions which may be enumerated explicitly. This will give a list which must include the enumerator polynomials of every real code. However, in general it will also include a number of ``fake'' enumerator polynomials. We cannot tell if a polynomial corresponds to a real code without additional information. This motivates our study of higher-weight enumerator polynomials, which are related to the higher genus partition functions.

\subsection{Lattices and CFTs from codes: Construction A}
\label{subsec:Lattices_from_Codes}

An error-correcting code can be viewed as a collection of points in $\F_2^n$. Let us imagine embedding this cube into $\mathbb{Z}^n$, and then requiring symmetry under translating any direction by two. The result is a lattice in $\mathbb{Z}^n$. This is known as Construction A, originally due to Leech and Sloane \cite{LeechSloane1971}. More precisely, the lattice $\Lambda$ associated to a code $\mathcal{C}$ by Construction A is defined as 
\begin{align}
    \Lambda(\mathcal{C}) = \left\{ v / \sqrt 2 \ | \ v \in \mathbb{Z}^n, \ v \equiv c \ (\text{mod } 2), \ c \in \mathcal{C} \right\}.
    \label{eq:def_lambda}
\end{align}
Now we may explain why we are concerned with doubly-even self-dual codes. First, consider a doubly-even code. The Hamming weight of all codewords must be divisible by four. If $v \equiv c \ (\text{mod } 2) $, and $c$ is a codeword whose weight is divisible by $4$, then $v^2$ is also divisible by $4$. The element of the lattice, $v / \sqrt 2$, therefore has a square-length that is divisible by $2$. A lattice where every vector has an even norm is, by definition, an \emph{even lattice}.

Next consider the lattice of the dual of code, $\Lambda(\mathcal{C}^\perp)$. Consider an element $\tilde v / \sqrt 2$ of this lattice, and take its inner product with an element $v / \sqrt 2$ of the lattice $\Lambda(\mathcal{C})$. 
\begin{align}
\begin{split}
        \frac{\tilde v}{\sqrt 2} \cdot \frac{v}{\sqrt 2} & = \frac{1}{2} (\tilde c + 2 \vec a) \cdot (c + 2 \vec b) 
    \\
    & =
    \frac{1}{2} \tilde c \cdot c + \vec a \cdot c + \vec b \cdot \tilde c + 2 \vec a \cdot  \vec b,
    \label{eq:dual_code_dual_lattice}
\end{split}
\end{align}
where in the first line, we have used the definition of an element of a code lattice, with $\vec a$ some element of $\mathbb{Z}^n$. The dual of a lattice is defined as the lattice of points with integer inner products with points in the original lattice. The inner product in (\ref{eq:dual_code_dual_lattice}) is an integer if and only if $\tilde c \cdot c$ is even. But this is the definition of the dual code $\mathcal{C}^\perp$. Therefore the lattice of the dual of a code is the same as the dual of the lattice of a code:
\begin{align}
    \Lambda(\mathcal{C}^\perp) = \Lambda(\mathcal{C})^* \, .
\end{align}
In summary, each doubly-even self-dual error-correcting code defines an even self-dual lattice via Construction A. These lattices are particulary interesting because they are the precisely the lattices which define conformal field theories. This idea was originally discussed by Narain \cite{Narain:1985jj, Narain:1986am} for the case of Lorentzian lattices. The idea is to consider a free CFT compactified on a lattice. Then requiring modular invariance of the partition function amounts to the requirement that the lattice is even and self-dual. 
In the present case, we are discussing Euclidean lattices. Compactifying $c$ free bosons on a $c$-dimensional lattice results in a chiral CFT with central charge $c$.

\subsubsection{Lattice theta-functions}
\label{subsec:latticethetasgenus1}

We will need one more element which will make the connection between codes and CFTs clearer -- the lattice theta function. On one hand, the lattice theta-function can be related to the partition function of the CFT corresponding to free bosons compactified on that lattice. On the other hand, for a lattice described by a code, the lattice theta function will be directly related to the enumerator polynomial. 
An integral self-dual lattice $\Lambda$ of rank $n=0\ \text{(mod $8$)}$, equipped with a Euclidean metric can be associated with a lattice theta series, or theta function, $\Theta_\Lambda$.
The (genus 1) theta-function of a lattice is defined by
\begin{align}\label{genus 1 lattice theta series}
    \Theta_\Lambda(\tau) = \sum_{v \in \Lambda} q^{v^2 / 2} \, , \qquad q = e^{2 \pi i \tau} \, .
\end{align}
The theta-function of a $c$-dimensional even self-dual lattice is related to the partition function of a CFT by
\begin{align}
    Z(\tau) = \frac{\Theta_\Lambda(\tau)}{\eta(\tau)^c} \, . 
\end{align}
Here $\eta(\tau)$ denotes the Dedekind eta function 
\begin{equation}
    \label{eq:dedekindeta}
    \eta(\tau)=e^{\frac{2i \pi \tau}{24}}\prod_{k=1}^\infty(1-e^{2\pi i k \tau}).
\end{equation}
Partition functions of 2d CFTs are required to be invariant under the modular transformations
\begin{align}
\label{eq:TandStransf}
    T: \quad \tau \to \tau + 1 \, , \qquad \qquad S: \quad \tau \to -\frac{1}{\tau} \, .
\end{align}

The theta-function for the dual lattice may be related to the the original theta-function by
\begin{align}
\label{eq:Lattice:dual}
    \Theta_{\Lambda^*}(\tau) = \mu (-i \tau)^{c / 2} \Theta(-1 / \tau) \, ,
\end{align}
where $\mu$ is a volume factor which is one for self-dual lattices. Using the fact that $\eta(-1/\tau) = \sqrt{-i \tau} \eta(\tau)$, it is clear that self-dual lattices correspond to partition functions which are invariant under $S$. If the lattice is even, then $\Theta(\tau) = \Theta(\tau + 1)$. However the Dedekind eta function picks up a phase under $T$ transformations: $\eta(\tau + 1) = \exp(2 \pi i / 24) \, \eta(\tau)$. This phase cancels when the number of eta functions is a multiple of 24. The result is that \emph{modular invariance of chiral CFTs requires that the central charge is divisible by 24}.

\subsubsection{Theta-functions and enumerator polynomials}

We are now ready to complete the correspondence between error-correcting codes and conformal field theories. The key element is the relation between the enumerator polynomial and the lattice theta-function (and thus the CFT partition function). The relation is
\begin{align}
    \Theta_{\Lambda(\mathcal{C})}(\tau) = W_\mathcal{C} (\theta_3(q^2), \theta_2(q^2) ),
    \label{eq:theta_enumerator_relation}
\end{align}
where $\theta_3$ and $\theta_2$ are two of Jacobi's theta functions. These are defined by\footnote{In Mathematica, $\texttt{EllipticTheta[}m\texttt{,0,}q\texttt{]}=\theta_m(q^2)$.}
\begin{align}
\label{eq:jacobithetas}
    \theta_1(q) &=  \sum_{n = -\infty}^\infty  (-1)^n q^{\frac{(n+ 1/2)^2}{2}}= 0 \, , 
    &  \theta_2(q) &= \sum_{n = -\infty}^\infty q^{\frac{(n+ 1/2)^2}{2}} \, , \\
    \theta_3(q) &= \sum_{n = -\infty}^\infty q^{\frac{n^2}{2}} \, , &  \theta_4(q) &= \sum_{n = -\infty}^\infty  (-1)^n q^{\frac{n^2}{2}}.
\end{align}
Essentially, \eqref{eq:theta_enumerator_relation} states that the theta function of a lattice corresponding to a code is computed by the enumerator polynomial of the code with the Jacobi $\theta$ functions as arguments. As we will be generalizing equation \eqref{eq:theta_enumerator_relation} to higher genus in section~\ref{sec:Higher Genus}, it will be useful to present its derivation. First, consider that each vector in the lattice lies in the equivalence class defined by a specific codeword. So we may rewrite the sum as the sum over all vectors as
\begin{align}
    \vec v = \frac{\vec c + 2 \vec a}{\sqrt 2}, \qquad \vec c\in \mathcal C,\quad \vec a\in \mathbb Z^n
\end{align}
So we can rewrite the theta function:
\begin{align}
\label{eq:thetaproof1}
\begin{split}
        \Theta(q) & = \sum_{\vec c, \vec a} q^{\frac{\vec{c}^2}{4}} q^{\vec c \cdot \vec a + \vec{a}^2}\\
    & = \sum_{\vec c} q^{\frac{\vec{c}^2}{4}} \left( \sum_{a_1} q^{c_1 a_1 + a_1^2}\right)\left( \sum_{a_2} q^{c_2 a_2 + a_2^2}\right)\cdots \left( \sum_{a_n} q^{c_n a_n + a_n^2}\right) ,
\end{split}
\end{align}
where we have expanded using the components $c_i$ and $a_i$ of the vectors $\vec c$ and $\vec a$. Now, $\vec c$ has $w(c)$ entries equal to $1$, and $n-w(c)$ entries equal to $0$, so
\begin{align}
\label{eq:thetaevaluatedgenus1}
\begin{split}
    \Theta(q) & = \sum_{\vec c} q^{\frac{w(c)}{4}} \left( \sum_j q^{j + j^2} \right)^{w(c)} \left( \sum_j q^{ j^2} \right)^{n - w(c)} \\
    & = \sum_{\vec c} \left( \sum_j q^{(j + 1/2)^2} \right)^{w(c)} \left( \sum_j q^{ j^2} \right)^{n - w(c)} \\
    & = \sum_{\vec c} \left( \theta_2(q^2) \right)^{w(c)}  \left( \theta_3(q^2) \right)^{n - w(c)} \\
    & = W_{\mathcal{C}} \left(  \theta_3(q^2),  \theta_2(q^2) \right),
\end{split}
\end{align}
which finishes the proof.

For convenience when we consider the higher genus case, let us introduce the following notation
\begin{equation}
\label{eq:definitionvarthetas}
    \vartheta_0(\tau):=\theta_3(q^2),\quad \vartheta_1(\tau):=\theta_2(q^2), \qquad q=e^{2\pi i \tau}.
\end{equation}
With this definition, we can write
\begin{equation}
    \Theta_{\Lambda(\mathcal C)}(\tau)=W_{\mathcal C}\left(\vartheta_0(\tau),\vartheta_1(\tau)\right),
\end{equation}
where the right-hand side is $W_{\mathcal C}$ evaluated at $x_A=\vartheta_A(\tau)$. Note that the standard modular transformations of the Jacobi theta functions imply that the $\vartheta_A(\tau)$ transform as
\begin{align}
    &T:\quad \vartheta_0(\tau)\mapsto \vartheta_0(\tau)\,, \quad \vartheta_1(\tau)\mapsto \vartheta_1(\tau)\,,\\
    &S:\quad \vartheta_0(\tau)\mapsto \sqrt{-i\tau}\frac{\vartheta_0(\tau)+\vartheta_1(\tau)}{\sqrt2}\,,\quad \vartheta_1(\tau)\mapsto \sqrt{-i\tau}\frac{\vartheta_0(\tau)-\vartheta_1(\tau)}{\sqrt2}
\end{align}
The $S$ transformation mimics the MacWilliams identity \eqref{eq:EP_s} and thus guarantees the transformation \eqref{eq:Lattice:dual} of the lattice theta function.

\section{Chiral CFTs at Higher Genus}
\label{sec:Higher Genus}

Having reviewed the relation between error-correcting codes and CFTs, we are now ready to develop the central points of this paper, which is how this correspondence works at higher genus. The results of this are summarized in table~\ref{tab:genus1highergenus}. At genus 2 and 3, this is relatively straightforward. After developing this correspondence, we will be able to show, in section~\ref{sec:enumeration}, how the modular invariance at genus 2 and 3 effectively constrain the space of possible enumerator polynomials, and therefore of code CFTs. At genus $g>3$, the situation is more complicated, as we will review in section~\ref{subsec:maths}.

\begin{table}
\centering
\caption{Summary of the differences between genus 1 and higher genus.}\label{tab:genus1highergenus}
{\renewcommand{\arraystretch}{1.75}
\begin{tabular}{>{\RaggedRight\arraybackslash}p{7cm}>{\RaggedRight\arraybackslash}p{7cm}}
Genus $g=1$ & General genus
\\\hline
Enumerator polynomial $W(x_0,x_1)$, & Higher-weight enumerator polynomial $W(x_0,\ldots,x_{2^g-1})$,
\\
Modular parameter $\tau$, & Period matrix $\Omega$,
\\
Modular group $\mathrm{PSL}(2,\mathbb Z)$ acting on $\{\IM\tau>0\}$, & Modular group $\mathrm{PSp}(2g,\mathbb Z)$ acting on the Siegel upper half-plane $\{\Omega=\Omega^T,\IM\Omega\succ0\}$,
\\
$\vartheta_A(\tau)$ expressible in terms of Jacobi theta functions $\theta_{2,3}(q^2)$,& $\vartheta_A(\Omega)$ expressible in terms of theta constants of second order characteristic $\thetaUD{\vek c_i/2}{\vek 0}(0,2 \Omega)$,
\\
Evaluation at $\vartheta_A$: $W(\vartheta_0(\tau),\vartheta_1(\tau))$, & Map $\Th:\ x_A\mapsto \vartheta_A(\Omega)$.
\\\hline
\end{tabular}
}
\end{table}

\subsection{Higher genus partition function from codes}

The central question answered by this paper is how the correspondence reviewed in the previous section can be generalized to higher genus Riemann surfaces. The higher genus partition function for a 2d CFT on a lattice can be expressed schematically as
\begin{align}
    \label{eq:PF_gen_g}
    Z^{(g)}(\Omega) =\frac{\hat Z^{(g)}(\Omega)}{\Phi_g(\Omega)} .
\end{align}
The partition function $Z^{(g)}$ depends on the period matrix $\Omega$, which describes the Riemann surface in an analogous way to $\tau$ in the genus 1 case. The numerator $\hat Z^{(g)}$ of the partition function is equal to the higher genus lattice theta function for theories defined by a self-dual even lattice $\Lambda$,
\begin{align}
    \hat Z^{(g)}(\Omega)= \Theta_{\Lambda}( \Omega) \, .
\end{align}
As we will discuss, this function has simple transformations under the higher genus modular group (it is a Siegel modular form of weight $c/2$). 

We will primarily be interested in the numerator, but let us make some comments about the denominator, $\Phi_g(\Omega)=F_g(\Omega)^c$, which corresponds to a sum over oscillator modes. 
At each genus, $F_g(\Omega)$ is universal and can in principle be evaluated by considering just one representative lattice CFT. At genus 1, $F_1(\tau)= \eta(\tau)$. At higher genus, no simple compact expression is known for the denominator. There exist some formal expressions \cite{McIntyre:2004xs} and at genus 2 a useful series expansion \cite{Gaberdiel:2010jf}. 
While these constructions give a denominator with the correct weight to cancel the weight of the modular form in the numerator, the resulting partition function picks up phases under the modular transformations that generalize the $T$ transformation in \eqref{eq:TandStransf}. At genus 1, one needs $c=24k$ for integer $k$ to cancel phases and render $Z^{(1)}$ completely modular invariant, and indeed, $\Delta_{12}=\eta(\tau)^{24}$ is a Siegel modular form of degree $12$, see section~\ref{subsec:Siegel}. For higher genus, it is not generally possible to find a nowhere vanishing denominator that is also a Siegel modular form,\footnote{At genus $2$, there is a nowhere vanishing modular form of degree 10: $\chi_{10}$, see section~\ref{subsec:Siegel}.} meaning that, in general, the partition function will always be defined up to phases.\footnote{If the chiral CFT is paired with its complex conjugate to give a full CFT, all phases picked up by modular transformations are canceled and the full CFT is modular invariant for any $c$. This is what happens, for instance, in the case of Narain CFTs \cite{Maloney:2020nni}.}

\subsubsection{Lattice theta functions at higher genus}
\label{subsec:gen_g_theta series}

We consider a compact Riemann surface $\Sigma$ with genus $g$ with $2g$ cycles $a_i,b_i$ where $i \in \{1,\ldots,g\}$. The canonical choice for the cycles is for their intersection numbers $\iota(,)$ to satisfy 
\begin{equation}\label{intersection normalization}
    \iota(a_i,a_j)=\iota(b_i,b_j)=0,\qquad \iota(a_i,b_j)=\delta_{ij}.
\end{equation}
A choice of $a$ cycles fixes the normalization of the $g$ holomorphic $1$ forms $\omega_j$ associated with the surface according to $\oint_{a_i} \omega_j = \delta_{ij}$. The $b$ cycles fix the matrix defined by 
\begin{equation}
    \oint_{b_i} \omega_j = \Omega_{ij}.
\end{equation}
$\Omega_{ij}$ is called the Riemann period matrix associated with $\Sigma$ and is a symmetric matrix with positive definite imaginary part: $\IM \Omega \succ 0$. 

The genus $g$ lattice theta series is defined as a sum over $g$-tuples of lattice vectors by 
\begin{align}
\label{eq:higher_lattice_theta}
    \Theta_\Lambda(\Omega)=\sum_{\vec v_1,\ldots, \vec v_g \in\Lambda}e^{ \pi i \vec v_i\cdot \vec v_j\Omega_{ij}}.
\end{align}
For instance, at genus 2 this gives
\begin{align}
    \Theta^{g=2}_\Lambda(\Omega)=\sum_{\vec v_1, \vec v_2 \in \Lambda} e^{\pi i (\vec v_1 \cdot \vec v_1 \Omega_{11} + 2 \vec v_1 \cdot \vec v_2 \Omega_{12} + \vec v_2 \cdot \vec v_2 \Omega_{22} )} ,
\end{align}
which can also be rewritten as
\begin{equation}
\label{eq:Thetagenus2form}
    \Theta^{g=2}_\Lambda(\Omega) = \sum_{\vec v_1, \vec v_2 \in \Lambda} q^{\frac{\vec v_1 \cdot \vec v_1 }{2}}  r^{ \vec v_1 \cdot \vec v_2} s^{\frac{ \vec v_2 \cdot \vec v_2 }{2}},
\end{equation}
with the the modular parameters $q,r,s$ are defined as
\begin{equation}
\label{eq:qrs}
    q=e^{2 \pi i \Omega_{11}}, \qquad r=e^{2 \pi i\Omega_{12}}, \qquad s=e^{2 \pi i\Omega_{22}}.
\end{equation}

\subsubsection{Higher-weight enumerator polynomials}

Above we have seen that Construction A can be used to define a lattice from a code, and that the lattice theta function can be computed from the code enumerator polynomial by replacing the code variables with theta constants according to
\begin{equation}
\Theta^{g=1}_{\Lambda(\mathcal{C})}(q) = W_{\mathcal{C}} \left(  \vartheta_0(\tau),\vartheta_1(\tau) \right).
\end{equation}
Now we would like to construct the analogous function at higher genus, which reproduces the higher genus lattice theta series from the code. The appropriate function is the \emph{higher-weight} enumerator polynomial and has been known in the math literature for some time \cite{MacWilliams1972}, but to our knowledge its connection to higher genus CFTs has never been pointed out. 

The higher-weight enumerator polynomial generalizes the weight 1 enumerator polynomial by comparing more than one codeword at a time. We will see later that this is related to the genus $g$ partition function. Specifically, the $g^{\text{th}}$ weight polynomial is defined by summing over $g$-tuples of codewords, and can be represented compactly as \cite{Oura2008}
\begin{align}
    W^{(g)}_{\mathcal{C}}(x_i) \ = \ \sum_{\mathbf{M} \in \mathcal{C}^g} \prod_{i = 1}^n x_{\text{row}_i(\bf{M})}.
    \label{eq:def_EP}
\end{align}
In this expression, the sum over all $g$-tuples of codewords is represented by the sum over all choices of the $n \times g$ matrix $\bf{M}$. Each column of $\bf{M}$ is a codeword, which is vector of length $n$ for an $n$-dimensional code. $\mathrm{row}_i(\mathbf M)$ denotes the $i$th row of $\mathbf M$. When writing $x_{\mathrm{row}_i}$, we interpret the index of $x$ as the binary representation of the index. Specifically $x_0=x_{[0\ldots00]}$, $x_1=x_{[0\ldots01]}$, $x_2=x_{[0\ldots10]}$, etc. We can see from this that the weight $g$ enumerator polynomial will be a homogeneous degree $n$ polynomial in the $2^g$ variables $x_0,x_1,\ldots x_{2^g-1}$.

For a specific case of this abstract definition, consider the $g = 2$ enumerator polynomial, sometimes called the biweight polynomial. It can be written as 
\begin{align}
    W^{(2)}_{\mathcal{C}}(x_0,x_1,x_2,x_3) = \sum_{c_1, \, c_2 \, \in \mathcal C} x_0^{n + c_1 \cdot c_2 - w(c_1) - w(c_2)} x_1^{w(c_2) - c_1 \cdot c_2} x_2^{w(c_1) - c_1 \cdot c_2}  x_3^{c_1\cdot c_2} \, .
    \label{eq:g2_EP}
\end{align}
We sum over all pairs of codewords. The exponents of the variables now involve the dot product of $c_1$ and $c_2$, as well as their individual weights. The result is a homogeneous degree $n$ polynomial of four variables. For example, the biweight polynomial of the $[8, 4, 4]$ Hamming code is given by
\begin{align}
\begin{split}
    W^{(2)}_{\text{Hamming}} \ &= \ x_0^8 + x_1^8 + x_2^8 + x_3^8 + 168 \, x_0^2 \,  x_1^2 \,  x_2^2 \,  x_3^2 \\
    & + 14 \, x_0^4 \, x_1^4 + 14 \, x_0^4 \,  x_2^4 + 14 \, x_0^4 \,  x_3^4 + 14 \, x_1^4 \,  x_2^4 + 14 \, x_1^4 \,  x_3^4+ 14 \, x_2^4 \,  x_3^4.
    \label{eq:hamming_g2}
\end{split}
\end{align}
In fact, we will see below that this is the unique degree eight genus two polynomial which satisfies some appropriate symmetries which we will derive in the next subsection.

Proving that this polynomial captures the theta function amounts to repeating the proof presented in \eqref{eq:thetaproof1} and \eqref{eq:thetaevaluatedgenus1} for our higher genus theta function. The key is to use the fact that each lattice vector is in the equivalence class of a codeword,
\begin{align}
    \vec v_i = \frac{\vec c_i + 2 \vec m_i}{\sqrt 2},
\end{align}
to rewrite the lattice theta function \eqref{eq:higher_lattice_theta} as 
\begin{align}
\begin{split}
        \Theta_{\Lambda(\mathcal{C})}(\Omega) \ &= \ \sum_{ \{\vec c_1, \ldots,\vec c_g \} \in \mathcal C}  \ \sum_{ \{ \vec m_1, \ldots, \vec m_g \} \in \mathbb{Z}^n} e^{ \frac{1}{2}\pi i   (\vec c_i + 2 \vec m_i)\cdot  (\vec c_j + 2 \vec m_j)  \Omega_{ij}}.
\end{split}
\end{align}
Now we want to consider the elements $\vec c_i$ entry-by-entry. So we define $c_{(i)k}$ to denote the $k^{th}$ entry of $\vec c_i$. Then the lattice theta function equals
\begin{align}
\begin{split}
        \Theta_\Lambda(\Omega) \ &=  \sum_{ \{\vec c_1, \ldots,\vec c_g \} \! \in \mathcal C^g} \ 
        \left(\sum_{ \{ m_{(1)1}, \ldots,  m_{(g)1} \} \in \mathbb{Z}^g} e^{ \frac{1}{2} \pi i   ( c_{(i)1} c_{(j)1} +2 c_{(i)1} m_{(j)1} +   2  m_{(i)1}c_{(j)1} + 4 m_{(i)1} m_{(j)1})  \Omega_{ij}} \right) \times \\
        & \qquad \qquad  \left(\sum_{ \{ m_{(1)2}, \ldots,  m_{(g)2} \} \in \mathbb{Z}^g} e^{ \frac{1}{2} \pi i   ( c_{(i)2} c_{(j)2} +2 c_{(i)2} m_{(j)1} +   2  m_{(i)2}c_{(j)2} + 4 m_{(i)2} m_{(j)2})  \Omega_{ij}} \right) \times\\
         & \qquad \cdots \left(\sum_{ \{ m_{(1)n}, \ldots,  m_{(g)n} \} \in \mathbb{Z}^g} e^{ \frac{1}{2} \pi i   ( c_{(i)n} c_{(j)n} +2 c_{(i)n} m_{(j)n} +   2 m_{(i)n} c_{(j)n} + 4 m_{(i)n} m_{(j)n})  \Omega_{ij}} \right)
\end{split}
\end{align}
Next we will define $\textbf{M} = \{ \vec c_1, \ldots,\vec c_g \} \in \mathcal{C}^g$. Then the $c_{(i)k}$ appearing in the $k^{th}$ term in parentheses form the $k^{th}$ row of $\textbf{M}$. Let us define a vector $2 \vek a_k = c_{(i)k}$. Furthermore, for any $k$, the range of $m_{(i)k}$ in the sum is the same, so let us define $\vek m = \vek m_{(i)k}$. With these replacements, the lattice theta function becomes
\begin{align}
\label{eq:thetagLambdaproduct}
        \Theta_\Lambda(\Omega) \ = \ \sum_{ \textbf{M} \in \mathcal C^g} \ 
        \left(\sum_{ \vek m \in \mathbb{Z}^g} e^{  \pi i  ( \vek a_1  + \vek m) \cdot (2\Omega) \cdot ( \vek a_1  + \vek m)} \right) \times \cdots \times  \left(\sum_{ \vek m \in \mathbb{Z}^g} e^{  \pi i  ( \vek a_n  + \vek m) \cdot (2\Omega) \cdot ( \vek a_n  + \vec m)} \right) .
\end{align}
At this point, we need to introduce the higher genus versions of the Jacobi theta functions. These are conventionally defined by (see \emph{e.g.}\ \cite{AlvarezGaume:1986es})
\begin{equation}
\label{eq:highergenustheta}
    \theta\begin{bmatrix}\vek a\\\vek b\end{bmatrix}(\vek z,\Omega)=\sum_{\vek m\in\Z^g}\exp\left( \pi i(\vek m+\vek a)\cdot\Omega\cdot(\vek m+\vek a)+2\pi i(\vek m+\vek a)\cdot(\vek z+\vek b)\right).
\end{equation}
We can see that each of the terms in parentheses in \eqref{eq:thetagLambdaproduct} is equal to a theta constant. So we find
\begin{align}
        \Theta_\Lambda(\Omega) \ &= \ \sum_{ \textbf{M} \in \mathcal C^g} \ 
        \left (\theta\begin{bmatrix}\vek a_1\\\vek 0 \end{bmatrix}(0,2 \Omega) \right) \times  \cdots \times  \left (\theta\begin{bmatrix}\vek a_n \\\vek 0 \end{bmatrix}(0,2 \Omega) \right).
\end{align}
Now recall that $ 2 \vek a_i = \vek c_i$ are the rows of $\textbf{M}$. So this formula is equal to the genus $g$ enumerator polynomial with the $x_i$ replaced by a corresponding theta expression. Following \cite{Oura2008}, we call this replacement the \emph{theta map}:
\begin{align}
\label{eq:thetamaphighergenus}
    \Th:\, x_{[c_1, c_2, \ldots]} \mapsto \thetaUD{\vek c/2}{\vek 0} (0,2 \Omega).
\end{align}
More precisely, the specific functions that appear on the right-hand side of \eqref{eq:thetamaphighergenus} are known as theta constants of second order characteristic \cite{Oura2008}. 
We will make more comments on the theta map in section \ref{sec:beyond}, where we will see that it is injective only for $g<3$ and surjective only for $g<4$ \cite{Oura2008b}. 

Now recall that $x_A = x_{[a_1, a_2, \ldots]}$ if $[a_1, a_2, \ldots]$ is the binary representation of $A$. Then it is convenient to introduce the following short form of the theta constants of second order characteristic,
\begin{align}
\label{eq:thetaconstantsdefn}
    \vartheta_A(\Omega) := \thetaUD{\vek a/2}{\vek 0}(0,2 \Omega) \, ,
\end{align}
allowing us to write the theta map as
\begin{align}
\label{eq:thetamapdef}
    \Th: x_A \mapsto \vartheta_A(\Omega) \, .
\end{align}
This leads to one of the central equations we will use in this paper:
\begin{align}
    \Theta_{\Lambda}(\Omega) = W^{(g)}_{\mathcal{C}}\left( \vartheta_0(\Omega), \,   \vartheta_1(\Omega) \, , \ldots ,  \vartheta_{2^g-1}(\Omega) \right).
\end{align}
In order to use this relationship to constrain code CFTs, we will need to determine the higher-weight MacWilliams identities obeyed by the enumerator polynomials. These transformations were already known in the math literature on error-correcting codes \cite{MacWilliams1972}, and the connection to modular forms was stressed in \cite{Runge1993, Runge1993b, Runge1995, Runge1996}. Here, we will review how to derive these relations, taking the viewpoint that they arise as a consequence of modular invariance of the higher genus partition function.

\subsection{Modular transformations at higher genus}

Two-dimensional chiral CFTs on a Riemann surface of genus $g$ enjoy more symmetries than just the modular transformations of a genus-one surface. The higher genus modular group is $\mathrm{Sp}(2g, \mathbb{Z})$, which arises from the fact that there are different ways of choosing the cycles satisfying \eqref{intersection normalization} on a given Riemann surface. Specifically, for an element
\begin{align}
     \begin{pmatrix}A&B\\C&D\end{pmatrix}\in  \mathrm{Sp}(2g,\mathbb{Z}) \, ,
\end{align}
the period matrix $\Omega_{ij}$ transforms as 
\begin{equation}
\label{eq:modularomega}
    \Omega\mapsto\tilde \Omega=(A\Omega+B)(C\Omega+D)^{-1}.
\end{equation}
These transformations act on the lattice theta function according to
\begin{equation}
\label{eq:lattice theta transformation}
    \Theta_\Lambda (\tilde \Omega) = \text{det} (C \Omega+D )^{\frac{n}{2}} \Theta_\Lambda\left(\Omega\right)  \, .
\end{equation}
The determinant factor in front of this is canceled by the transformation of the denominator factor to ensure that the partition factor is invariant (up to phase). 

In \cite{Stanek1963}, an explicit basis for the generators of $\mathrm{Sp}(2g, \mathbb{Z})$ is given. 
It is generated by three matrices for $g = 2, 3$ and only two matrices for $g>3$. We give the generators for $g = 2$ and $g = 3$, which we call $T_g$, $R_g$, and $D_g$. These will allow us to determine the transformations of the code variables $x_i$. For $g = 2$, we have
\begin{align}
        T_{g = 2} = \begin{pmatrix}
        1 & 0 & 1 & 0\\
        0 & 1 & 0 & 0\\
        0 & 0 & 1 & 0\\
        0 & 0 & 0 & 1
        \end{pmatrix} \, , \ \ R_{g = 2} = \begin{pmatrix}
        1 & 0 & 0 & 0\\
        1 & 1 & 0 & 0\\
        0 & 0 & 1 & -1\\
        0 & 0 & 0 & 1
        \end{pmatrix} \, , \ \ D_{g = 2} = \begin{pmatrix}
        0 & 1 & 0 & 0\\
        0 & 0 & -1 & 0\\
        0 & 0 & 0 & 1\\
        1 & 0 & 0 & 0
        \end{pmatrix}.
\end{align}
For $g = 3$, the generators are $6 \times 6$ matrices:
\begin{align}
    \begin{split}
        &T_{g = 3} = \begin{pmatrix}
        1 & 0 & 0 & 1 & 0 & 0\\
        0 & 1 & 0 & 0 & 0 & 0\\
        0 & 0 & 1 & 0 & 0 & 0\\
        0 & 0 & 0 & 1 & 0 & 0\\
        0 & 0 & 0 & 0 & 1 & 0\\
        0 & 0 & 0 & 0 & 0 & 1
        \end{pmatrix} \, , \qquad R_{g = 3} = \begin{pmatrix}
        1 & 0 & 0 & 0 & 0 & 0\\
        1 & 1 & 0 & 0 & 0 & 0\\
        0 & 0 & 1 & 0 & 0 & 0\\
        0 & 0 & 0 & 1 & -1 & 0\\
        0 & 0 & 0 & 0 & 1 & 0\\
        0 & 0 & 0 & 0 & 0 & 1
        \end{pmatrix} \, , \\
        &\qquad \qquad \qquad  D_{g = 3} = \begin{pmatrix}
        0 & 1 & 0 & 1 & 0 & 0\\
        0 & 0 & 1 & 0 & 0 & 0\\
        0 & 0 & 0 & -1 & 0 & 0\\
        0 & 0 & 0 & 0 & 1 & 0\\
        0 & 0 & 0 & 0 & 0 & 1\\
        1 & 0 & 0 & 0 & 0 & 0
        \end{pmatrix}.
    \end{split}
\end{align}

These generators determine how the transformations act on the period matrix $\Omega$ through \eqref{eq:modularomega}. This, in turn, determines the transformation of the theta constants. The derivation requires use of the Poisson resummation formula. As it is rather tedious, we will merely record the answers. 
\begin{align}
\begin{split}
    T_{g = 2} &:  \quad \vartheta_0(\Omega)   \mapsto   \vartheta_0(\Omega), \quad \, 
    \vartheta_1(\Omega) \mapsto   \vartheta_1(\Omega), \quad  \, 
    \vartheta_2(\Omega) \mapsto    i \vartheta_2(\Omega), \quad  \, 
    \vartheta_3(\Omega) \mapsto i  \vartheta_3(\Omega) ,\\
    R_{g = 2}& :  \quad \vartheta_0(\Omega)   \mapsto  \vartheta_0(\Omega), \quad  \,  
    \vartheta_1(\Omega) \mapsto  \vartheta_3(\Omega), \quad  \, 
    \vartheta_2(\Omega) \mapsto   \vartheta_2(\Omega), \quad  \, 
    \vartheta_3(\Omega) \mapsto  \vartheta_1(\Omega), \\
    D_{g = 2} &:  \quad \vartheta_0(\Omega)   \mapsto  \sqrt{-i \Omega_{11}} \ \frac{\vartheta_0(\Omega) + \vartheta_2(\Omega)}{\sqrt 2}, \qquad 
    \vartheta_1(\Omega) \mapsto  \sqrt{-i \Omega_{11}} \ \frac{\vartheta_0(\Omega) - \vartheta_2(\Omega)}{\sqrt 2}, \\
    &   \qquad  \vartheta_2(\Omega) \mapsto   \sqrt{-i \Omega_{11}} \ \frac{\vartheta_1(\Omega) + \vartheta_3(\Omega)}{\sqrt 2}, \qquad 
    \vartheta_3(\Omega) \mapsto \sqrt{-i \Omega_{11}} \ \frac{\vartheta_1(\Omega) - \vartheta_3(\Omega)}{\sqrt 2} \, .
    \label{eq:genus2thetasymmetries}
\end{split}
\end{align}
It is important that $\det( C \Omega + D)$ is equal to 1 for $T_{g =2}$ and $R_{g = 2}$ and $\Omega_{11}$ for $D_{g = 2}$. This ensures
\begin{align}
    D_{g=2}: \qquad  W^{(2)}_{\mathcal{C}}\left( \vartheta_0(\Omega),\ldots \right) \mapsto \det (C \Omega+D )^{\frac{n}{2}} W^{(2)}_{\mathcal{C}}\left( \frac{\vartheta_0(\Omega) + \vartheta_2(\Omega)}{\sqrt 2} ,\ldots \right)
\end{align}
and likewise for any other $\mathrm{Sp}(4, \mathbb{Z})$ transformation. Therefore we see that the transformation of the theta function \eqref{eq:lattice theta transformation} requires that the enumerator polynomial satisfy the \emph{higher-weight MacWilliams identities}:
\begin{align}
\begin{split}
    T_{g = 2} &:  \qquad x_0   \mapsto   x_0, \qquad  \ \
    x_1 \mapsto   x_1, \qquad \ \ 
    x_2 \mapsto    i x_2, \qquad \ \ \,
    x_3 \mapsto i  x_3 ,\\
    R_{g = 2}& :  \qquad x_0   \mapsto  x_0, \qquad \quad \,  
    x_1 \mapsto  x_3, \qquad \ \ \, 
    x_2 \mapsto   x_2, \qquad \ \ \, 
    x_3 \mapsto  x_1, \\
    D_{g = 2} &:  \qquad x_0   \mapsto  \frac{x_0 + x_2}{\sqrt 2}, \quad 
    x_1 \mapsto  \frac{x_0 - x_2}{\sqrt 2}, \quad 
    x_2 \mapsto   \frac{x_1 + x_3}{\sqrt 2}, \quad 
    x_3 \mapsto  \frac{x_1 - x_3}{\sqrt 2} \, ,
    \label{eq:genus2symmetries}
\end{split}
\end{align}
For genus 3 the analysis is the same. We skip the transformations of the theta constants and simply present
 \begin{align}
 \begin{split}
     T_{g = 3} &:  \qquad x_4 \to i x_4 \, , \quad x_5 \to i x_5 \, , \quad x_6 \to i x_6 \, , \quad x_7 \to i x_7 \, ,  \\
     R_{g = 3}& : \qquad x_2 \to  x_6 \, , \quad x_3 \to  x_7 \, , \quad x_6 \to  x_2 \, , \quad x_7 \to  x_3 \, ,  \\
     D_{g = 3} &:  \qquad x_0   \to  \frac{x_0 + x_4}{\sqrt 2} \, , \quad  x_1   \to  \frac{x_0 - x_4}{\sqrt 2} \, , \quad  x_2   \to  \frac{x_1 + x_5}{\sqrt 2} \, , \quad x_3   \to  \frac{x_1 - x_5}{\sqrt 2} \, , \\
     & \ \ \, \qquad  x_4   \to  \frac{x_2 + x_6}{\sqrt 2} \, , \quad x_5   \to  \frac{x_2 - x_6}{\sqrt 2} \, , \quad x_6   \to  \frac{x_3 + x_7}{\sqrt 2} \, , \quad x_7   \to  \frac{x_3 - x_7}{\sqrt 2} \, .
     \label{eq:genus3symmetries}
 \end{split}
 \end{align}
These transformations will be crucial to the algorithm we describe in section~\ref{sec:enumeration}. For genus 2 they were already known \cite{MacWilliams1972}. We do not know if they have been written down in the code literature for genus 3. They effectively express a very complicated set of transformations, \emph{i.e.}\ the higher genus modular transformations of the partition function, in terms of a small set of linear transformations. We use the term ``invariant polynomials'' to refer to those polynomials which are unchanged by these transformations. 

\subsection{Factorization limits}
\label{sec:factorization}

\begin{figure}[h]%
    \centering
\includegraphics{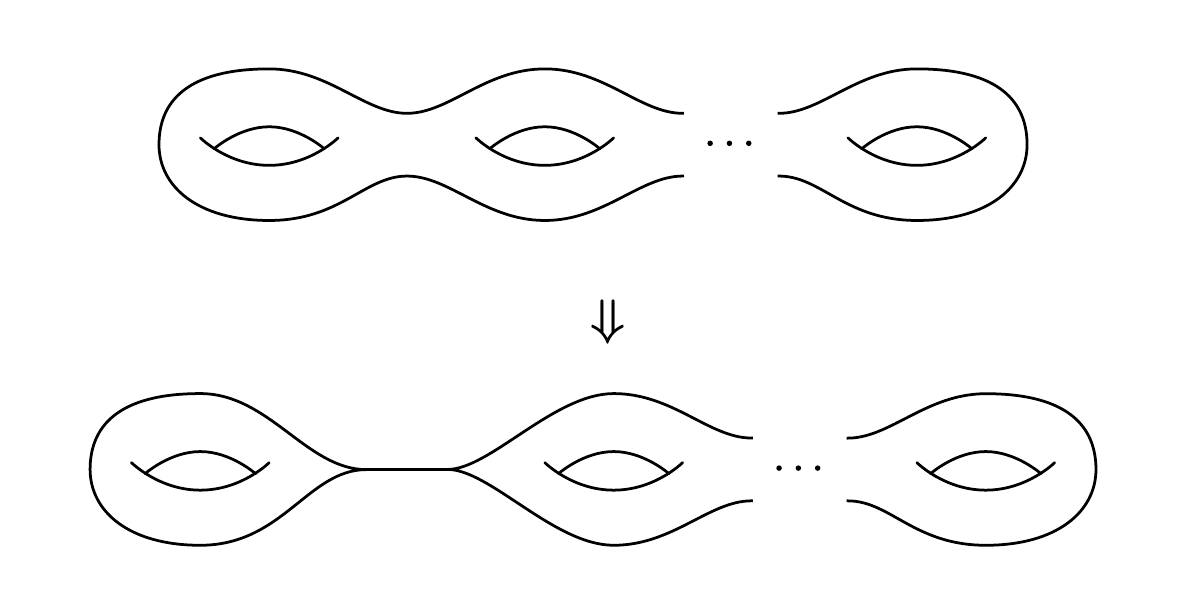}
    \caption{Factorization  of a genus $g$ Riemann surface. }
    \label{fig:genusg-fig}%
\end{figure}

Further constraints may be imposed on the higher genus partition function based on the limit where the Riemann surface becomes singular \cite{Friedan:1986ua}. For a genus $g$ Riemann surface $\Sigma_g$ we may smoothly deform $\Sigma_g$ so that it consists of two Riemann surfaces of genus $h$ and $g-h$, connected by a long, infinitely thin tube. In this limit, the genus $g$ partition function must behave as the product of a genus $h$ partition function, and a genus $g-h$ partition function. We refer to these deformations as ``factorization limits''.
The case of $h=1$ is depicted in figure~\ref{fig:genusg-fig}. 

In the factorization limit, the period matrix becomes block-diagonal. For instance, for $h=1$,
\begin{equation}
    \Omega\to \tau\oplus \Omega',
\end{equation}
where $\tau=\Omega_{11}$ and $\Omega'_{ij}=\Omega_{i+1,j+1}$ (\emph{i.e.}\ we have $\Omega_{1J}=\Omega_{J1}=0$). From the CFT perspective, in the factorizatin limit only the identity operator flows across the thin tube, and the partition function is expected to factorize into a product of the lower genus partition functions. Equivalently, for a lattice theta function,
\begin{equation}
\label{eq:factorizationTheta}
    \Theta_\Lambda^{(g)}(\Omega)\to \Theta_\Lambda^{(g)}(\tau \oplus \Omega')=\Theta_\Lambda^{(1)}(\tau)\Theta_\Lambda^{(g-1)}(\Omega').
\end{equation}
The case of arbitrary $h$ is completely analogous. 

We shall see that factorization imposes strong constraints on which enumerator polynomials can correspond to CFT partition functions. There are a number of higher-weight polynomials which are modular invariant but which do not factorize properly. This allows us to eliminate them from the list of possible partition functions. Let us see how the factorization property~\eqref{eq:factorizationTheta} acts on the level of the polynomial variables $x_i$. First recall the binary form of the genus $g$ polynomial variables,
\begin{equation}
    x_{[i_{g-1}i_{g-2}\cdots i_{0}]}^{(g)}\ \longleftrightarrow \ x_i^{(g)}, \qquad i=\sum_{n=0}^{g-1}i_n2^n.
\end{equation}
Then \eqref{eq:factorizationTheta} is consistent with the factorization map
\begin{equation}
    \label{eq:factorizationX}
    x_{[i_{g-1}i_{g-2}\cdots i_{0}]}^{(g)}\mapsto  x^{(1)}_{i_{g-1}}\, y^{(g-1)}_{[i_{g-2}\cdots i_0]},
\end{equation}
or, equivalently
\begin{equation}
 \label{eq:factorizationXalt}
    x_i\mapsto\begin{cases}
     x_0\,y_i,& 0\leqslant i<2^{g-1},
    \\
     x_1\,y_{i-2^{g-1}}, & 2^{g-1}\leqslant i<2^g.
    \end{cases}
\end{equation}
Here $x_i$ and $y_i$ refer to the polynomial variables of the two resulting Riemann surfaces. It can be seen that these definitions are consistent with the factorization property of code enumerator polynomials,
\begin{equation}
\label{eq:factorizationolys}
W_{\mathcal C}^{(g)}(x_i)\mapsto W_{\mathcal C}^{(1)}(x_i)W_{\mathcal C}^{(g-1)}(y_i),
\end{equation}
under \eqref{eq:factorizationX}.

\begin{figure}[h]%
    \centering

\includegraphics{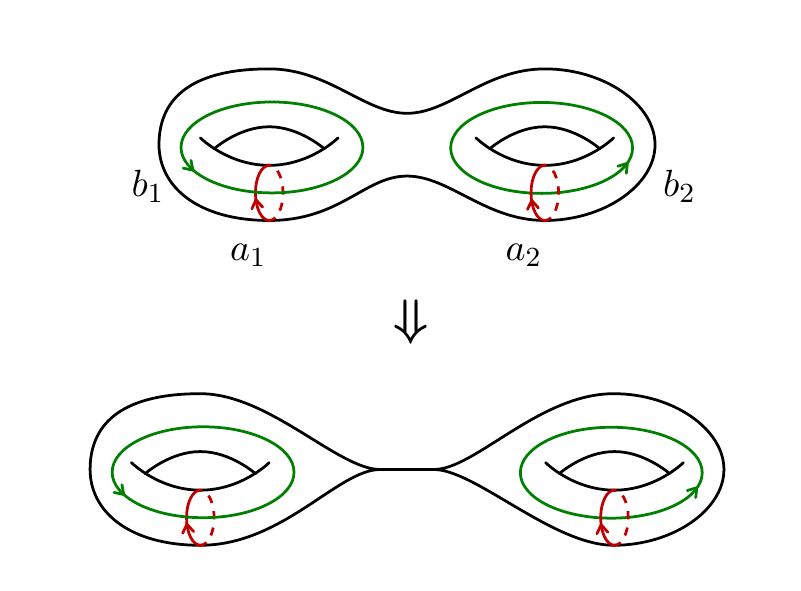}
    \caption{Factorization of a genus 2 Riemann surface. The $a$ cycles are depicted in red, $b$ cycles in green. Under the factorization, the one-form $\omega_2$ dual to $a_2$ vanishes on the left torus. Hence $\Omega_{12}=\oint_{b_1}\omega_2=0$, so that \ $r=e^{2\pi i \Omega_{12}}\to 1$.}
    \label{fig:genus2-fig}%
\end{figure}

\paragraph{Genus 2 example:} It is easy to see that this is correct requirement with the simple example of genus 2. On the genus $2$ Riemann surface one can choose $a$ cycles and $b$ cycles as indicated in figure~\ref{fig:genus2-fig}, and introduce the variables $q=e^{2\pi i\Omega_{11}}$, $r=e^{2\pi i \Omega_{12}}$ and $s=e^{2\pi i\Omega_{22}}$ according to \eqref{eq:qrs}.
In the degeneration limit, $q$ and $s$ reduce to the modular parameters $e^{2\pi i \tau}$ and $e^{2\pi i\tau'}$ of the respective tori.
Moreover, the one-form $\omega_2$ dual to the $a_2$ cycle on the right torus becomes approximately zero on the left torus (and vice versa), showing that $\Omega_{12}\to0$, or equivalently, the $r\to1$ limit.\footnote{There are subleading corrections to factorization \cite{Tuite:1999id}, which are conveniently expressed in plumbing fixture coordinates giving a recipe for stitching together two tori with modular parameter $\tau_1,\tau_2$ via a tube of radius $\epsilon$. This gives $r=1+O(\epsilon)$. For our purposes we shall always consider the complete factorization $r=1$.}
For the lattice theta function in this example, we have (see \eqref{eq:Thetagenus2form})
\begin{equation}
    \Theta^{g=2}_\Lambda(\Omega) = \sum_{\vec v_1, \vec v_2 \in \Lambda} q^{\frac{\vec v_1 \cdot \vec v_1 }{2}}  r^{ \vec v_1 \cdot \vec v_2} s^{\frac{ \vec v_2 \cdot \vec v_2 }{2}}.
\end{equation}
In the limit $r\to1$, the resulting lattice theta function then breaks into the product of two sums, 
\begin{equation}
    \Theta^{g=2}_\Lambda(\Omega_{12} = 0) = \sum_{\vec v_1 \in \Lambda} q^{\frac{\vec v_1 \cdot \vec v_1 }{2}} \sum_{ \vec v_2 \in \Lambda} s^{\frac{ \vec v_2 \cdot \vec v_2 }{2}},
\end{equation}
which are each equal to a genus 1 lattice theta function. For the polynomial variables, it is easy to see from the definition of $\theta$
\begin{align}
    x_{[0,0]} = \theta\begin{bmatrix}0\\0\\0\\0\end{bmatrix}(0,2\Omega) \xrightarrow{\Omega_{12} \to 0} \left(\theta\begin{bmatrix}0\\0\end{bmatrix}(0,2\Omega_{11}) \right) \left( \theta\begin{bmatrix}0\\0\end{bmatrix}(0,2\Omega_{22}) \right) = x_0 \, y_0 \, ,
\end{align}
etc.

It is also sometimes convenient to consider the Siegel map $\Phi$. This maps the polynomial variables according to
\begin{equation}
\label{eq:Siegel}
    \Phi(x_i^{(g)})=\begin{cases}
    x_g^{(g-1)},& 0\leqslant i<2^{g-1},
    \\
    0, & 2^{g-1}\leqslant i<2^g.
    \end{cases}
\end{equation}
The Siegel $\Phi$ map is equivalent to taking first the factorization $Z^{(g)}\to Z^{(1)}Z^{(g-1)}$ and then taking the limit $\tau\to i\infty$, which sets $ x_0=1$, $ x_1=0$.

\subsection[Genus $g>3$ and Siegel modular forms]{Genus $\boldsymbol{g>3}$ and Siegel modular forms}
    \label{subsec:maths}
    
    In most of this paper we are interested in the case where $g = 2$ or $g = 3$, mainly because these are the cases where the enumerator polynomials are small enough to make explicit calculations tractable. Here we would like to broaden our discussion to any genus, which will require some additional mathematical formalism. The reader primarily interested in our results at genus 2 and 3 may proceed directly to section~\ref{sec:enumeration}. 
    
    In particular, we are interested in properties of the evaluation map $\Th$, which sends polynomial variables (functions of $x_i$) to theta expressions (functions of $\Omega$). The $\Th$ map is a ring homomorphism between invariant polynomials and modular covariant functions in the Siegel upper half plane
    \begin{align}
        \mathcal H_g=\{\Omega=\Omega^T,\IM \Omega\succ0\} \, .
    \end{align}
    Such covariant forms are known in the mathematics literature as Siegel modular forms. In section~\ref{subsec:Siegel} we review the most important facts about Siegel modular forms in order to explain the properties of the $\Th$ map at higher genus. Starting at genus 3, the $\Th$ map ceases to be one-to-one; there exist a degree $16$ polynomial $j_8$ which maps to $0$. This implies a non-trivial relation among the genus-3 theta functions. More complications arise at even higher genus, which we will discuss in section~\ref{sec:beyond}
    
\subsubsection{Siegel modular forms}
\label{subsec:Siegel}

A Siegel modular form (for the group $\mathrm{Sp}(2g,\Z)$) of degree $k$ is a function $f_k:\mathcal H_g \to\C$ which transforms covariantly with weight $k$. This is to say that under the transformation
\begin{equation}
    \Omega\mapsto \Omega'=(A\Omega+B)(C\Omega+D)^{-1}, \quad \begin{pmatrix}
    A&B\\C&D
    \end{pmatrix}\in \mathrm{Sp}(2g,\Z) \, ,
\end{equation}
$f_k$ transforms as
\begin{equation}
    f_k(\Omega')=\det(C\Omega+D)^kf_k(\Omega).
\end{equation}
The space of Siegel modular forms at genus $g$ defines a ring graded by degree, $M_g=\bigoplus M_g^k$. If we let $R_g$ be the ring of invariant polynomials, \emph{i.e.}\ those satisfying the generalized MacWilliams identities, the map $\Th:R_g\to M_g$ defined by \eqref{eq:thetamaphighergenus} becomes a ring homomorphism compatible with the grading. In fact, we have the commuting diagram \cite{Oura2008}
\begin{equation}
\label{eq:commutingdiagramRM}
\begin{tikzcd}
R_g^{2k} \arrow[r, "\Th"]  \arrow[d," \Phi"] & M^k_g \arrow[d,"\Phi"] 
\\
R^{2k}_{g-1} \arrow[r,"\Th"] & M^k_{g-1} ,
\end{tikzcd}
\end{equation}
where $\Phi$ is the Siegel map \eqref{eq:Siegel}.
Here $R_g$ has the obvious grading by degree.
The ring $M_g$ of modular forms (of even degree) is known for $g\leqslant3$, and has the following generators
\begin{align}
    & g=1 & & G_4,\ G_6,
    \label{eq:generatorsgenus1}
    \\
    & g=2 && E_4,\ E_6,\ \chi_{10},\ \chi_{12},
    \label{eq:generatorsgenus2}
    \\
    \label{eq:generatorsgenus3}
    & g=3 && \begin{matrix*}[l]\alpha_4, \ \alpha_6,\ \alpha_{10},\, \alpha_{12},\, \alpha'_{12},\, \beta_{14},\, \alpha_{16},\, \beta_{16},\, \chi_{18},\, \alpha_{18},\, \\\alpha_{20},\, \gamma_{20},\, \beta_{22},\, \beta'_{22},\, \alpha_{24},\,\gamma_{24},\,\gamma_{26},\, \chi_{28},\, \alpha_{30},\end{matrix*}
\end{align}
where the indices denote the degree of the respective generator.
The result at genus $2$ is due to Igusa \cite{Igusa1962}. At genus $3$, an overcomplete list of $34$ generators was given in \cite{Tsuyumine1986}, and was recently reduced to the minimal set of $19$ generators listed in \eqref{eq:generatorsgenus3}  \cite{Lercier2019}.\footnote{The normalization of the genus $3$ generators differ between \cite{Tsuyumine1986} and \cite{Lercier2019}; we will not need the normalization of the genus 3 modular forms.} At genus 1 and 2, the ring is generated by the holomorphic Eisenstein series: $G_k$, $k=4,6$ (genus $1$) and $E_k$, $k=4,6,10,12$ (genus $2$). See appendix~\ref{eq:holoeisenstein} for relations between the generators at genus 2.

 Since we have $c$ divisible by $8$, only Siegel modular forms of degree divisible by four will enter our discussion. At genus 1, this subring can be generated by $G_4$ and any degree $12$ modular form. A convenient choice of the latter is $\Delta_{12}=\frac1{1728}(G_4^3-G_6^2)$. In the next section, we will find the following expressions for these forms:\footnote{In the literature, the following compact expressions are also common
\begin{equation}
\label{eq:G4alt}
    G_4=\frac12\left(\theta_2(q)^8+\theta_3(q)^8+\theta_4(q)^8\right),\qquad 
    \Delta_{12}=\eta(\tau)^{24}= \frac1{256}\theta_3(q)^8\theta_2(q)^8\theta_4(q)^8,
\end{equation}
where the Jacobi theta functions $\theta_2$, $\theta_3$ and $\theta_4$ have arguments $q$ and not $q^2$ as in \eqref{eq:G4D12dir} (see the definition \eqref{eq:definitionvarthetas}).
}
\begin{align}
\label{eq:G4D12dir}
\begin{split}
    G_4(\tau)&=\vartheta_0(\tau)^8+\vartheta_1(\tau)^8+14\vartheta_0(\tau)^4\vartheta_1(\tau)^4,
    \\
    \Delta_{12}&=\frac1{16}\vartheta_0(\tau)^4\vartheta_1(\tau)^4\left(\vartheta_0(\tau)^4-\vartheta_1(\tau)^4\right)^4.
    \end{split}
\end{align}
Likewise, at genus 2, it is convenient to introduce the notation $\psi_{12}=\frac1{1728}(E_4^3-E_6^2)$.

The map $\Th:R_g\to M_g$ is one-to-one for genus $g=1$ and $g=2$. However, at genus $3$, there is a degree $16$ polynomial $j_8^{(3)}(x_0,\ldots x_7)$ which maps to a non-trivial relation among the genus $3$ theta constants. It turns out that this is the only such relation at genus $3$ \cite{Runge1993}. In summary, for $g\leqslant3$ we have
\begin{equation}
\label{eq:Ringisomorphisms}
    M_1\cong R_1, \qquad M_2\cong R_2, \qquad M_3 \cong R_3/\langle j_8^{(3)}\rangle,
\end{equation}
where $\langle j_8^{(3)}\rangle$ denotes the ideal generated by $j_8^{(3)}$. So we see that at $g=3$, $\Th$ fails to be injective, but the quotient in \eqref{eq:Ringisomorphisms} makes the situation easy to deal with. The situation at higher genus is further complicated by additional relations, and by the fact that $\Th$ is no longer surjective either, as we discuss in section~\ref{sec:beyond}.

\paragraph{Siegel modular forms under factorization}
In the factorization limit, defined in section~\ref{sec:factorization}, genus $g$ Siegel modular forms factor into products of lower-genus forms. For the purposes of this paper, we will only need to consider the factorization of genus 2 Siegel modular forms, which is given by (with $q=e^{2 \pi i \tau_1},s=e^{2 \pi i \tau_2}$):
\begin{align}\label{Seigel forms factorization genus 2}
\begin{split}
    E_4(q, r, s) &\longrightarrow G_4(q) \, G_4(s) ,\\ 
    E_6(q, r, s) &\longrightarrow G_6(q) \, G_6(s) ,\\ 
    \chi_{10}(q, r, s) &\longrightarrow 0 ,\\
    \chi_{12}(q, r, s) &\longrightarrow \Delta_{12}(q) \, \Delta_{12}(s) ,\\ 
    \psi_{12}(q, r, s) &\longrightarrow G_4(q)^3 \, \Delta_{12}(s) +  \Delta_{12}(q) \, G_4(s)^3 - 1728 \Delta_{12}(q) \,  \Delta_{12}(s).
\end{split}
\end{align}

\paragraph{Siegel modular forms under degeneration limits}
The Siegel $\Phi$ operator relates modular forms at different genus according to equation \eqref{eq:Siegel}. For the forms of degree up to 12, it acts as follows:
\begin{align}\label{eq:Siegel-forms-normalization}
\begin{split}
    \alpha_4&\mapsto E_4\mapsto G_4\mapsto 1,
    \\
    \alpha_6&\mapsto E_6\mapsto G_6\mapsto 1,
    \\
    \alpha_{10}&\mapsto \chi_{10}\mapsto0,
    \\
    \alpha_{12}&\mapsto \chi_{12}\mapsto0,
    \\
    \alpha'_{12}&\mapsto 0.
\end{split}
\end{align}
Forms that map to zero under $\Phi$ are called cusp forms. $\chi_{10}$, $\chi_{12}$ and $\alpha'_{12}$ are cusp forms. At genus $1$, the modular discriminant $\Delta_{12}$ is a cusp form.

\subsubsection{Beyond Genus 3}
\label{sec:beyond}

The method employed in this paper is suitable for genus $g\leqslant3$. Apart from the polynomial $j_8(x_i)$ at genus $3$, which maps onto a nontrivial relation among the theta functions, there is a one-to-one map between enumerator polynomals, Siegel modular forms, and CFT partition functions:
\begin{equation}
    W^{(g)}(x_i) \text{ (mod $j_8$)} \longleftrightarrow W^{(g)}(\vartheta_i(\Omega)) \longleftrightarrow Z^{(g)}(\Omega).
\end{equation}
Moreover, for genus $g\leqslant3$ the ring $M_g$ of genus $g$ Siegel modular forms (for $\mathrm{Sp}(2g,\Z)$) has been fully characterized, and one can search for partition functions in this space.
However, as summarized in table~\ref{tab:genus3highergenus}, starting from genus $4$ there are complications to each of these statements.
\begin{table}
\centering
\caption{Summary of the differences between genus $g\leqslant3$ and higher genus.}\label{tab:genus3highergenus}
{\renewcommand{\arraystretch}{1.75}
\begin{tabular}{>{\RaggedRight\arraybackslash}p{7cm}>{\RaggedRight\arraybackslash}p{7cm}}
Genus $g\leqslant3$ & Genus $g\geqslant4$
\\\hline
$\Th:R_g\to M_g$ surjective (one-to-one for $g\leqslant2$)
&
$\Th:R_g\to M_g$ neither surjective nor injective
\\
Ring $M_g$ of Siegel modular forms known & Ring $M_g$ unknown
\\
Moduili space $\mathcal M_g$ of Riemann surfaces dense inside moduli space $
\mathcal A_g$ of ppavs & Locus of $\mathcal M_g$ inside $\mathcal A_g$ is non-trivial, and not known for $g>4$
\\
\hline
\end{tabular}
}
\end{table}

Consider first the ring homomorphism $\Th:R_g\to M_g$ mapping invariant polynomials to Siegel modular forms. At genus 1 and 2 it is completely invertible. At genus 3 it is surjective but injectivity fails; the kernel is generated by $ j_8$, a degree 16 polynomial, see \eqref{eq:Ringisomorphisms}. At higher genus there exist more relations like $\Th(j^{(3)}_8)=0$. For instance, at genus 4, the first element of the kernel of $\Th$ is a degree 24 polynomial \cite{Freitag2001}, and there are additional relations, one at degree 28 and five at degree $32$ \cite{Oura2008}. (It is not known if additional relations exist at higher degree.) Starting from genus 4, moreover, it has been shown that the map $\Th$ is not even surjective \cite{Oura2008b},\footnote{We thank R Salvati Manni for pointing out this reference.} meaning that there are Siegel modular forms that do not descend from enumerator polynomials. For instance, as shown already in \cite{SalvatiManni1986}, at genus 5, the unique degree 6 Siegel modular form (which maps to $\alpha_6\mapsto E_6 \mapsto G_6$ under successive applications of the Siegel $\Phi$ operator) has no pre-image in $R_5$ -- no polynomial maps to it under $\Th$. Note, however, that there is a polynomial which maps to the genus 5 counterpart of $G_4$, as expected on the grounds that each code has an enumerator polynomial at every genus.\footnote{To see how this is possible, consider the problem of writing an alternative type of enumerator polynomials at genus $1$ in terms of the variables $a=\theta_3(q)$, $b=\theta_2(q)$ (\emph{i.e.}\ in terms of Jacobi theta functions with argument $q$ and not $q^2$). Then one finds from \eqref{eq:G4alt}, and the Jacobi relation $\theta_2(q)^4 - \theta_3(q)^4 + \theta_4(q)^4 = 0$, that $E_4=a^8+b^8-a^4b^4$, $\Delta_{12}=\frac1{256}a^8b^8(a^4-b^4)^2$, while $E_6=\sqrt{E_4^3-1728\Delta_{12}}$ is not a polynomial.} While non-surjectivity of $\Th$ does not prevent us from finding code CFTs using our method, it means that our method will not be able to find other CFTs whose high-genus partition function derives from Siegel modular forms but not from any enumerator polynomial.

Next, consider the fact that the ring of Siegel modular forms is unknown at higher genus. As reviewed above, the complete set of relations for at genus $3$ was only recently determined \cite{Lercier2019}, although the generators have been known for a while \cite{Tsuyumine1986}. At genus $4$ the complete set of generators is not even known. Combined with non-surjectivity of the map $\Th$, the lack of a complete set of generators poses another obstacle of extending our program beyond code CFTs at high genus.

Finally, consider the last obstacle, namely the construction of a partition function given a Siegel modular form $f\in M_g$. For concreteness, focus on the numerator of the partition function, $\hat Z\in \lambda^{c/2}(\mathcal M_g)$.\footnote{Here we stressed that the numerator of the partition function is not a function on $\mathcal M_g$, but instead a section on $\lambda^{c/2}(\mathcal M_g)$, where $\lambda(\mathcal M_g)$ denotes the determinant line bundle on $\mathcal M_g$. This in turn descends from the determinant line bundle on $\mathcal A_g$, see for instance comments in \cite{Putman2012}.} The relation to the Siegel upper half plane $\mathcal H_g$, on which the Siegel modular forms are functions, is as follows. Consider first the quotient space
\begin{equation}
     \mathcal A_g=\mathcal H_g/\mathrm{Sp}(2g,\Z),
\end{equation}
denoted the moduli space of principally polarized abelian varieties (ppavs). This space generalizes the fundamental region $-\frac12<\RE\tau<\frac12,\ |\tau|>1$ to higher genus. By modular covariance, Siegel modular forms $f\in M_g$ of weight $k$ are mapped to sections of the $k$th power of the determinant line bundle on $\mathcal A_g$. By the Torelli theorem, there is an embedding $\mathcal M_g\to\mathcal A_g$ mapping a Riemann surface to its Jacobian.\footnote{Concretely, the Jacobian of a Riemann surface with period matrix $\Omega$ is the complex torus given by $\C^g/(\Z^g+\Omega\Z^g)$}
For $g\leqslant 3$, the moduli space $\mathcal M_g$ of genus $g$ Riemann surfaces is dense in $\mathcal A_g$, meaning that almost every point in $\mathcal A_g$ represents a genus $g$ Riemann surface. However, as the dimension formulas
\begin{equation}
    \dim \mathcal M_g=\begin{cases}1&g=1,\\3g-3&g>1,\end{cases}\quad \qquad \dim \mathcal A_g=\frac12g(g+1)
\end{equation}
show, for $g\geqslant4$ the embedding of $\mathcal M_g$ inside $\mathcal A_g$ has a non-zero co-dimension. At genus $4$, with $\dim \mathcal M_4=9$, $\dim \mathcal A_4=10$, the locus of (the closure of) $\mathcal M_4$ inside $\mathcal A_4$ is known, and given by the vanishing of a specific degree eight modular form $J_8^{(4)}$:
\begin{equation}
    \overline{\Jac \mathcal M_4}=\{J^{(4)}_8=0\}\subset\mathcal A_4.
\end{equation}
At higher genus, the embedding is not known. The modular form $J_8^{(4)}$ is called the Schottky form, and is related to the polynomial $j_8^{(3)}$, as we will discuss in detail below.

There are two consequences of $\overline{\Jac \mathcal M_g}\neq \mathcal A_g$ for $g\geqslant4$. Firstly, the existence of non-zero Siegel modular forms $f$ which vanish on the moduli space, means that the map $f\mapsto \hat Z$ is not injective. More importantly, there may be sections $\hat Z$ of $\lambda^{c/2}(\mathcal M_g)$ (\emph{i.e.}\ numerators of partition function) that do not lift to sections of $ \lambda^{c/2}(\mathcal A_g)$ at sufficiently high genus. This means that when considering candidate CFTs with enumerator polynomial at $g\leqslant g_{\mathrm{max}}$ and which lack enumerator polynomial at $g>g_{\mathrm{max}}$, if $g_{\mathrm{max}}\geqslant3$ we cannot rule out the existence of a CFT corresponding to this set of enumerator polynomials.

\paragraph{The Schottky form and the codes associated to $\boldsymbol{e_8 \oplus e_8}$ and $\boldsymbol{d_{16}^+}$}

Let us now consider in more detail an example of the consequences of the considerations discussed here. This is Milnor's example \cite{Milnor} of isospectral lattices mentioned in the introduction, which is based on a result by Witt \cite{Witt1941}. Consider the codes associated via Construction A to the lattices $e_8\oplus e_8$ and $d_{16}^+$, which are the root lattices of $E_8\times E_8$ and $\mathrm{spin}(32)/\mathbb Z_2$ \cite{Witt1941}.\footnote{The code associated with the $e_8$ lattice is in fact the Hamming $[8,4,4]$ code, the unique even self-dual binary code at $c=8$.} In fact, $e_8=d_8^+$ and $d_{16}^+$ are the first two elements of a general class of lattices $d_{8k}^+$ with the following description (see \emph{e.g.}\ \cite{Conway1997}). Let $\Lambda_0=\{(v_1,\ldots v_{8k})\in \mathbb Z^{8k}\,|\,v_1+\ldots+v_{8k}=0\ (\mathrm{mod}\, 2)\}$. Then the lattice for $d_{8k}^+$ is given by
\begin{equation}
\label{eq:Lambdad8kplus}
    \Lambda_{d_{8k}^+}=\Lambda_0\cup\left(\tfrac12(1,1,\ldots,1)+\Lambda_0\right).
\end{equation}
In the sum defining the lattice theta function, the restriction to $v_1+\ldots+v_{8k}=0\ (\mathrm{mod}\, 2)$ can be implemented by $1\to\frac{1+(-1)^{v_1+\ldots+v_{8k}}}{2}$, and using the definition \eqref{eq:highergenustheta} of the higher genus theta functions we find that\footnote{Note that the argument here is $\Omega$ and not $2\Omega$ as in \eqref{eq:thetamaphighergenus}.} 
\begin{equation}
\label{eq:thetaford8kplus}
    \Theta_{d_{8k}^+}=\frac1{2^{g}}\sum_{\mathbf A}\theta[\mathbf A](\Omega)^{8k}.
\end{equation}
Here the sum is over all $2^{g-1}(2^g+1)$ vectors $\mathbf A\in\{\binom00,\binom{1/2}0,\binom0{1/2},\binom{1/2}{1/2}\}^g$ that give a non-zero theta constant.\footnote{At genus $g=1$, this list contains the three elements $A=\binom00,\,\binom{1/2}0$, and $\binom0{1/2}$, corresponding to the Jacobi theta functions $\theta_3(q),\,\theta_2(q)$ and $\theta_4(q)$ respectively, see \eqref{eq:jacobithetas}.}

Define $j_8^{(g)}$ to be the difference between the enumerator polynomials of the codes corresponding to the lattices $e_8\oplus e_8$ and $d_{16}^+$,
\begin{equation}
    j_8^{(g)}=(W^{(g)}_{e_8})^2-W^{(g)}_{d_{16}^+}.
\end{equation}
For the case of genus 3, we will recover an expression for $j_8^{(3)}$ in the next section, see \eqref{eq:j83implicit}. The polynomial $j_8^{(g)}$ evaluates under $\Th$ to the Schottky form $J_8^{(g)}$:
\begin{equation}
\Th(j_8^{(g)})=    J_8^{(g)}:=\frac1{2^g}\bigg(\bigg(\sum_{\mathbf A}\theta[\mathbf A](\Omega)^8\bigg)^2-2^g\sum_{\mathbf A}\theta[\mathbf A](\Omega)^{16}\bigg).
\end{equation}
It was first written down at genus $4$ in a different format by Schottky in 1888 \cite{Schottky1888} and was later found by Igusa \cite{Igusa1981} to be proportional to the expression above. 
As advertised in the previous section, at genus 4 the Schottky form has the important property that its vanishing solves the \emph{Schottky problem}, that is to determine the locus of $\mathcal M_g$ inside $\mathcal A_g$. 

Now return to the partition functions corresponding to the codes $e_8\oplus e_8$ and $d_{16}^+$. For genus $1$ and $2$, their enumerator polynomials agree: $W^{(g)}_{e_8\oplus e_8}=W^{(2)}_{d_{16}^+}$, $g=1,2$. At genus $3$, their enumerator polynomials are different, $W^{(3)}_{d_{16}^+}=W^{(3)}_{e_8\oplus e_8}-j_8^{(3)}$, but the associated Siegel modular forms are equal: $\Th(W^{(3)}_{d_{16}^+})=\Th(W^{(3)}_{e_8\oplus e_8})$. At genus $4$ the associated Siegel modular forms are different, but coincide on $\mathcal M_4$, so that $\hat Z^{(4)}_{d_{16}^+}=\hat Z^{(4)}_{e_8\oplus e_8}\in \lambda^8(\mathcal M_4)$.\footnote{Again, $\hat Z$ refers to the numerator of the partition function, see \eqref{eq:PF_gen_g}.} Finally, at genus $5$, it was shown in \cite{Grushevsky:2008zp} that at genus $5$ the partition functions in fact do differ,
\begin{equation}
    \hat Z^{(g)}_{d_{16}^+}=\hat Z^{(g)}_{e_8\oplus e_8}, \quad g\leqslant4, \qquad \hat Z^{(g)}_{d_{16}^+}\neq\hat Z^{(g)}_{e_8\oplus e_8}, \quad g\geqslant 5.
\end{equation}
An alternative proof of this fact was given in \cite{Gaberdiel:2009rd}. In conclusion, the example we have considered shows that at $c=16$ there are two chiral conformal field theories, corresponding to the codes/lattices $e_8\oplus e_8$ and $d_{16}^+$, that share partition functions for $g\leqslant4$ and are only distinguished at genus 5.

In fact, the non-vanishing of $J_8$ on $\mathcal M_5$ was found in \cite{Grushevsky:2008zp} with the motivation of computing the chiral superstring measure, and implies the non-vanishing of the cosmological constant at genus 5 in type II and heterotic superstring theory. This follows from an all-genus ansatz of the superstring measure proposed in \cite{Grushevsky:2008zm} based on results at lower genus \cite{DHoker:2001jaf,DHoker:2004fcs,Cacciatori:2008ay}. This ansatz, when summed over even spin structures (equivalent to the summation variables in \eqref{eq:Lambdad8kplus}), becomes proportional to $J_8^{(g)}$ (see \cite{Morozov:2008wz} for an overview).\footnote{An alternative ansatz was proposed in \cite{Oura2010}, also leading to non-vanishing cosmological constant at genus 5. The two ans\"atze can in principle be combined to give a vanishing genus 5 cosmological constant, but this approach runs into difficulties at genus 6, see \cite{Morozov:2008wz,Dunin-Barkowski:2009uak,Dunin-Barkowski:2012ial}. See also \cite{Matone:2010yv,Matone:2012wy} for a relation to the bosonic string theory measure.}

\section{Counting Higher Genus Enumerator Polynomials}
\label{sec:enumeration}

In section~\ref{sec:Higher Genus}, we examined how the correspondence between classical error-correcting codes and 2d chiral CFTs extends to higher genus. Now we would like demonstrate the utility of this correspondence by explicitly showing how the higher genus modular symmetries, combined with the required factorization properties, drastically shrinks the space of allowed enumerator polynomials. 

We aim to present the algorithm used in this section in a fairly self-contained way. However we will not include many details, such as the action of the modular group and the factorization limits, which are discussed in section~\ref{sec:Higher Genus}.

\subsection{Method}

In practice, our method is very simple. There are four steps:
\begin{enumerate}
    \item write down the most general genus $g$ polynomial,
    \item reduce undetermined coefficients by imposing the symmetries,
    \item list solutions where all coefficients are positive integers,
    \item eliminate all solutions which don't factorize.
\end{enumerate}

The logic behind these steps has been considered at length in the previous section, so let us just remind the reader of a few important points. Then we will turn to some specific examples, which should make the procedure clear.

In the first step, the polynomial we start with depends on both the genus $g$ and the dimension $c$ of the code. Specifically, it will be the most general homogeneous degree $c$ polynomial of $2^g$ variables. Recall the definition of the enumerator polynomial of a code: the coefficients of each term represent degeneracies of codewords. True enumerator polynomials must have positive integer coefficients, thereby motivating step 3.

The second step amounts to imposing modular invariance. This is a strict constraint on the polynomials which can possibly correspond to codes. The size of the modular group grows, leading to stricter constraints. However, the size of the polynomials in step 1 also increases quickly with the genus, so the number of solutions in step 2 will increase quickly with the genus. It is only by requiring the correct factorization limits, which we do in step 4, that we find a reduction in the number of potential codes. 

\subsection[Example: enumerating $c = 24$ polynomials]{Example: enumerating $\boldsymbol{c = 24}$ polynomials}

Let us now show how this works for the specific example of 24-dimensional codes. It was found by Conway (unpublished, see note in \cite{Pless1975}) that there are 9 doubly-even self-dual codes in 24 dimensions.

We will see that our algorithm restricts the number of potential enumerator polynomials. At genus 1, there are 190 possible polynomials which satisfy steps 1, 2, and 3. At genus 2, we will have an explosion in the number of possible polynomials which satisfy 1, 2 and 3. However only a tiny subset consisting of 29 polynomials, will properly factorize into genus 1 polynomials. This effectively rules out $190 - 29 = 161$ of the genus 1 polynomials. They cannot correspond to codes because if they did, they would lead to a factorizing genus 2 polynomial.  Repeating the procedure at genus 3 leads to a further reduction from 29 to 21 possible polynomials, as we will see. 

\subsubsection{Genus 1}

The genus 1, $c = 24$ partition function is an order-24 polynomial of $x_0$ and $x_1$. We start by writing the most general polynomial with one for the leading coefficient (which is required because the identity is always a codeword). 
\begin{align}
    P^{(1)}_{\text{gen}} = x_0^{24} + a_1 x_0^{23} x_1 + a_2 x_0^{22} x_1^2 + a_3 x_0^{21} x_1^3  +\ldots.
\end{align}
Next, we recall the invariances of this polynomial

\begin{align}
    x_0 \mapsto \frac{x_0 + x_1}{\sqrt 2} \, , \qquad x_1 \mapsto \frac{x_0 - x_1}{\sqrt 2} \, , 
\end{align}
and
\begin{align}
    x_1 \mapsto i x_1 \, .
\end{align}
Imposing these conditions places a number of constraints on the allowed values for $a_i$. The resulting polynomial therefore depends on far fewer coefficients:
\begin{align}
\begin{split}
    &P^{(1)}_{\text{inv}} = x_0^{24} +  a_{4} \,  x_0^{20} x_1^{4} +(759-4 a_{4}) x_0^{16} x_1^{8} + (2576 + 6 a_{4})x_0^{12} x_1^{12} \\
    & \hspace{40mm} + (759-4 a_{4}) x_0^{8} x_1^{16} + a_{4} \,  x_0^{4} x_1^{20} + x_1^{24}.
\end{split}
\end{align}
Now that we have determined the most general invariant polynomial, we must impose that all of the coefficients are positive integers. It is easy to see that this is the case when $a_{4}$ is an integer satisfying
\begin{align}
    0 \leq a_4 \leq 189 \, .
\end{align}
So we find that there are 190 possible genus 1 polynomials. Any possible ECC must have one of these enumerator polynomials, because no other polynomial satisfies all of the necessary conditions. Note that in the case of genus 1, there is no step 4.

Also note that we have only a single undetermined coefficient. This reflects the fact that the ring of genus $1$ Siegel modular forms has only two independent elements at degree 12: $G_4^3$ and $G_6^2$. This gives one degree of freedom once the leading coefficient is set to one. We shall find a similar story for genus 2 and genus 3.

To relate this result to the physical spectrum of the theory, we can write the partition function,
\begin{align}
    Z^{(1)}_{c = 24} =\frac{ P^{(1)}_{\text{inv}}}{\eta(\tau)^{24}},
\end{align}
and expand in small $q$ to find\footnote{Recall that the genus 1 partition function expands as $q^{-\frac{c}{24}}$ times an expansion in state multiplicities.}
\begin{align}
    Z^{(1)}_{c = 24}(\tau) = \frac{1}{q} \left( 1 + (72+ 16 a_4) q + 196884 q^2 + 21493760 q^3 + \ldots\right). 
\end{align}
From this expansion we can see that the number of conserved currents $N_{\text{currents}} = 72 + 16 a_4$. The genus 1 allowed polynomials, therefore, could have between 72 and 3,096 currents.

\subsubsection{Genus 2}

The algorithm for genus 2 is largely the same for the first three steps. First we have 
\begin{align}
    P^{(2)}_{\text{gen}} = x_0^{24} + a_{1,0,0} x_0^{23} x_1 + a_{0,1,0} x_0^{23} x_2 + a_{0,0,1} x_0^{23} x_3 + a_{2,0,0} x_0^{22} x_1^2 + \ldots.
\end{align}
The result is a 2,925-term polynomial in $x_0,x_1,x_2,x_3$. Next we impose the symmetries \eqref{eq:genus2symmetries}. The expression for $P^2_{\text{inv}}$ is very long, but it has a shorter set of unique coefficients, which must be positive integers. The result is
\begin{align}
\begin{split}
    a_{4,0,0} \ &\geqslant \ 0 \, , \\
    759-4 a_{4,0,0} \  & \geqslant \ 0 \, , \\
    2576 + 6 a_{4,0,0} \ & \geqslant \ 0 \, , \\
    2 a_{2,2,2} + 5 a_{4,0,0}\  & \geqslant \ 0 \, , \\
    -2 a_{2,2,2} + 250 a_{4,0,0}\  & \geqslant \ 0 \, , \\
    22770 + 36  a_{2,2,2} -540 a_{4,0,0}\  & \geqslant \ 0 \, , \\
    a_{2,2,2} \  & \geqslant \ 0 \, , \\
    -12 a_{2,2,2} + 480 a_{4,0,0}\  & \geqslant \ 0 \, , \\
    22 a_{2,2,2} + 2112 a_{4,0,0}\  & \geqslant \ 0 \, , \\
    340032 - 52  a_{2,2,2} - 2432 a_{4,0,0}\  & \geqslant \ 0 \, , \\
    212520 + 76  a_{2,2,2} -3480 a_{4,0,0}\  & \geqslant \ 0 \, , \\
    1275120 + 36  a_{2,2,2} -1530 a_{4,0,0}\  & \geqslant \ 0 \, , \\
    4080384 - 8  a_{2,2,2} + 29952 a_{4,0,0}\  & \geqslant \ 0 \, . \\
\end{split}
\end{align}
This list of unique coefficients implies some redundant inequalities, but we have include the entire set for completeness. Using Mathematica, we can easily extract the set of integer solutions to these inequalities. The result is a set of 135,065 solutions. Note also that there are two undetermined parameters after imposing $a_{0,0,0} = 1$. This is again a reflection that there are three independent modular forms at genus 2: $E_4^3$, $E_6^2$ and $\chi_{12}$ (see section~\ref{subsec:Siegel}).

Here we can demonstrate how step 4, the factorization limit, can be imposed to further restrict the number of polynomials. It is important to note the this will provide a restriction on the \emph{genus 1} polynomials as well: we will rule out all genus 2 polynomials which don't factorize properly, and we will rule out all genus 1 polynomials which don't arise from factorizing a genus 2 polynomial. The logic here is simple: a real code must have a genus 2 polynomial, and it must factorize into the square of that code's genus 1 polynomial.

Recall how the factorization limit acts on our genus 2 polynomials:\footnote{Technically we could represent the polynomials on each torus with different variables, \emph{e.g.} $x_0 \to x_0 y_0$, etc. In principle this could lead to stronger constraints on the allowed polynomials, but interestingly we find no difference in these procedures.}
\begin{align}
    x_0 \to x_0^2, \quad x_1 \to x_0 x_1, \quad  x_2 \to x_0 x_1, \quad x_3 \to x_1^2 \, .
\end{align}
So we merely need to check each of our 135,065 genus 2 polynomials. If the factorization limit turns it into the square of a genus 1 polynomial, then it is allowed.

Remarkably, there are only 29 such polynomials on our entire list! Each of these polynomials is the square of one of our 190 genus 1 polynomials. Therefore we have shown that only 29 of our genus 2 polynomials can possibly come from codes, and only 29 of our genus 1 polynomials can possibly come from codes. See table~\ref{tab:c=24} for a summary.

\begin{figure}[t]%
    \centering
    \subfloat[\centering]{{\includegraphics[width=7cm]{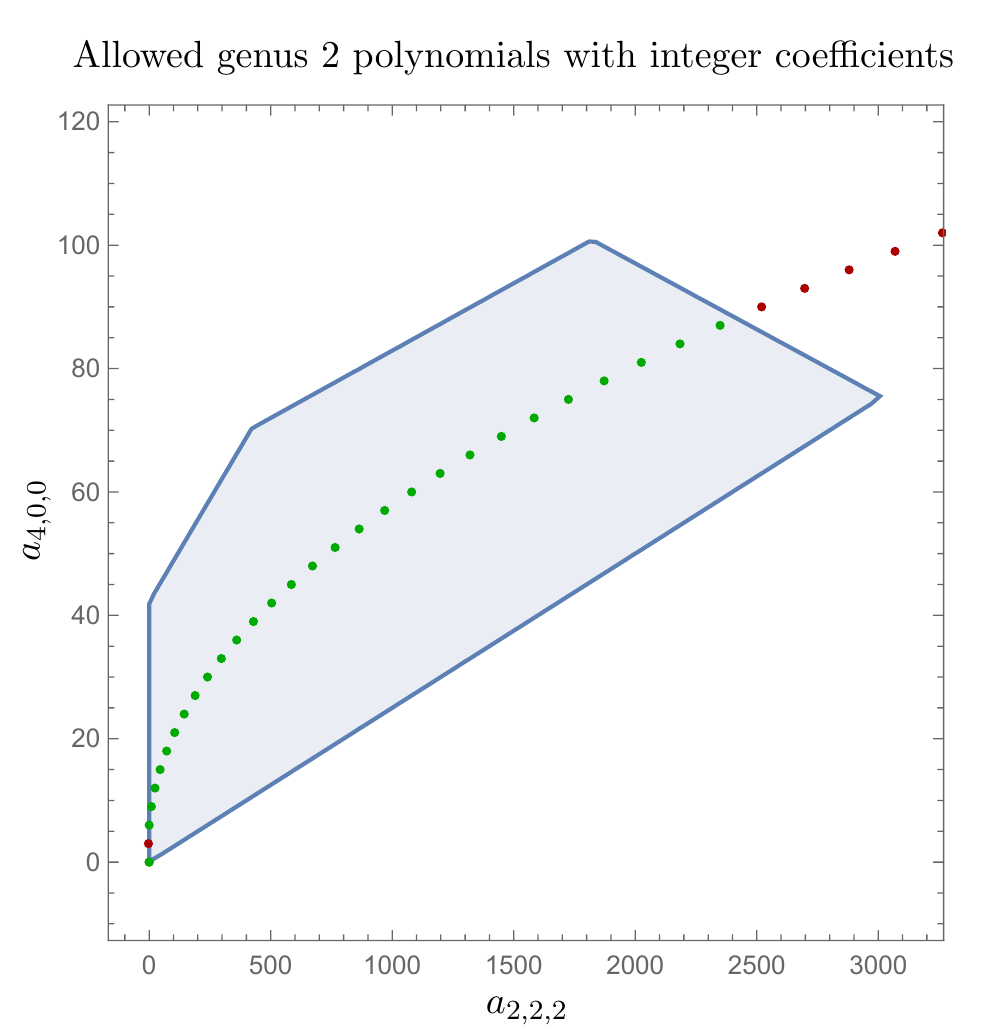} }}%
    \qquad
    \subfloat[\centering ]{{\includegraphics[width=7cm]{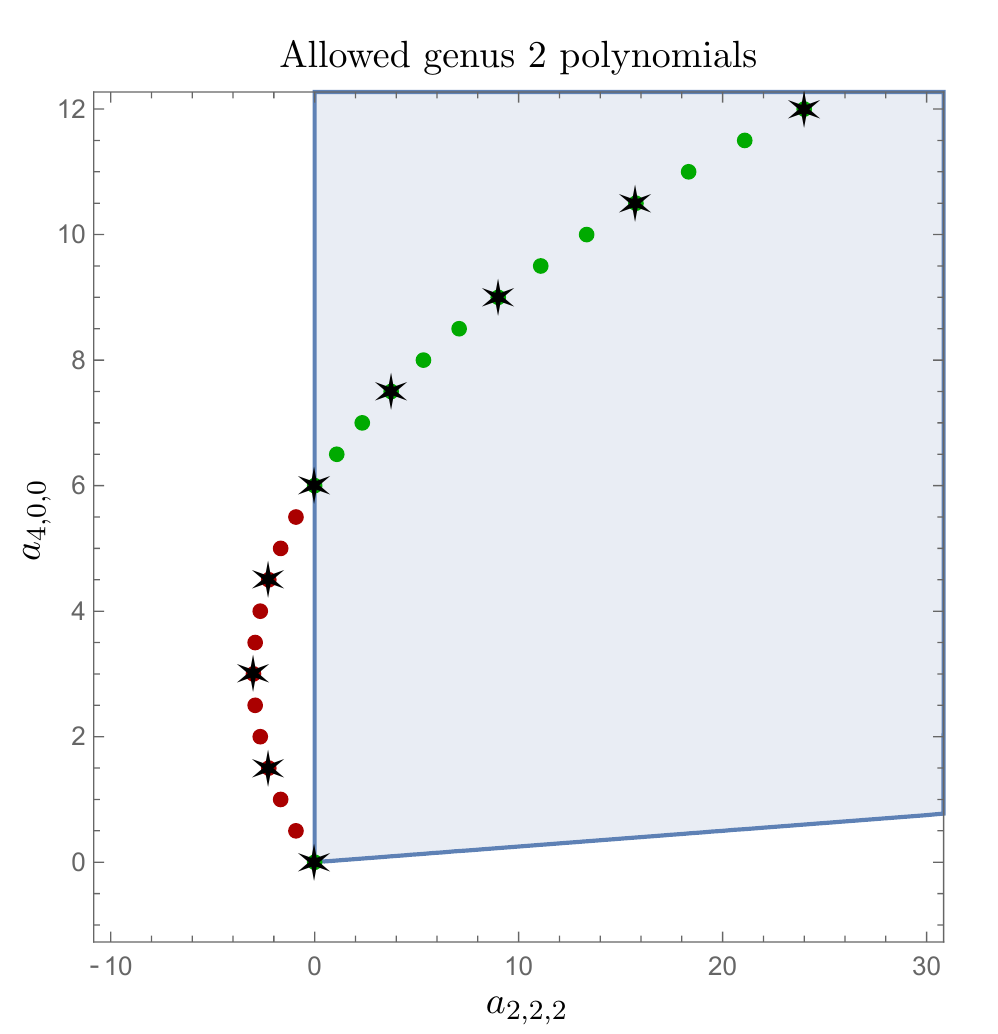} }}%
    \caption{Allowed polytope for the genus 2 coefficients. Green dots represent polynomials which lie inside the bounds, and red outside. Plot (a) shows all 29 allowed genus 2 polynomials with integer coefficients. In (b), we zoomed into the bottom left, and allowed for half-integer coefficients in $P^1_{\text{inv}}$ allowing a larger number of solutions. Black stars represent lattices. We can see the $k = 4$, $k = 5$, and $k = 6$ lattices outside the allowed region. 
    }%
    \label{fig:allowed g2}%
\end{figure}

For genus 2 at $c=24$, the set of constraints defines a two-dimensional polytope, and it is interesting to consider the way that the allowed polynomials sit inside it. This is displayed in figure~\ref{fig:allowed g2}. We see that a single genus 1 polynomial is excluded near the origin, and then a large number of polynomials are excluded in the upper range for the $a_{2,2,2}$. It is also interesting to note that $a_{4,0,0}$ satisfies $759 \geq 4 a_{4,0,0} \geq 0$, the same bound that we found on $a_4$ at genus 1. However the rest of the constraints eventually give stronger constraints on $a_{4,0,0}$, meaning that if we project figure~\ref{fig:allowed g2} (a) to the $y$ axis, and identify $a_{4,0,0} \sim a_4$, we find that the space of allowed genus 2 polynomials lies inside the space of allowed genus 1 polynomials. We believe this pattern will persist to higher $g$ and $c$. A final interesting feature is that the real theories display a quadratic relationship between $a_{2,2,2}$ and $a_{4,0,0}$. This appears to be related to the fact that, in equation (4.19) of \cite{Keller:2017iql}, there are two undetermined coefficients $b_2$ and $b_3$ appearing in the genus 2 partition function which are linear and quadratic in the number of currents $N_c$, respectively. It would be interesting to see if any version of this pattern persists -- already at $c = 32$, it is known that the undetermined coefficients can depend on the number of currents \emph{and} the OPE coefficients of low-lying states \cite{Keller:2017iql}.

\subsubsection{Genus 3}

The procedure for genus 3 follows the same idea as genus 1 and 2, however some tricks are required to deal with the large number of terms in the polynomial $P^{(3)}_{\mathrm{gen}}$ and the large number of solutions, \emph{i.e.}\ polynomials of the form $P^{(3)}_{\mathrm{inv}}$ with positive coefficients. 

First, we do not start with the most general polynomial; instead we eliminate from the beginning a large number of terms which could never appear (analogous to how terms with odd powers could never appear at genus 1, due to the $x_1 \to i x_1$ transformation). The precise procedure for doing this is laid out in \cite{Runge1996} in the discussion on ``admissible polynomials.''

Then we can impose the symmetries on this smaller polynomial to construct $P^{(3)}_{\mathrm{inv}}$. We find that it depends on four independent parameters. This should reflect  five independent modular forms at this order. However there are actually only four! This is due to the existence of a ``theta relation,'' discussed in section~\ref{sec:beyond}. This is a non-trivial combination of theta functions which equals zero, \emph{i.e.}\ an element of the kernel of the theta-map. The conclusion is that, at genus 3, the space of polynomials is larger than the space of modular forms.

Another trick we use is that we do not enumerate every solution. In general each coefficient of $P^{(3)}_{\mathrm{inv}}$ must be positive, giving an inequality. For genus 3 (or genus 2, at large $c$), it becomes impossible to enumerate every solution. However it is possible to determine which genus 1 polynomials can arise as the factorization limit of genus 3 polynomials without determining the full set of genus 3 polynomials. 

Our upgraded approach is thus to search through each of the 29 genus 1 polynomials which are allowed by genus 2 factorization, and determine if it can also arise from genus 3 factorization. Specifically, we look at the factorization limit of $P^{(3)}_{\mathrm{inv}}$ and then we solve for the undetermined $a$-coefficients for each of the genus 1 polynomials. In every case, we find solutions, but for some of the genus 1 polynomials, all of the solutions result in disallowed (negative coefficient) genus 3 polynomials. In this way, we are able to rule out more genus 1 polynomials from our list.

The final result from this procedure is given in table~\ref{tab:c=24}. From our original list of 190 polynomials, we first reduce to 29 genus 2 polynomials (this number was already known in \cite{Runge1996}). Then we find that demanding the existence of a consistent genus 3 polynomial further reduces this to a set of 21 polynomials. As far as we know, we are the first to count this number. It is easy to see from table~\ref{tab:c=24} that each of the polynomials which corresponds to a real code is allowed by our procedure.

It is possible to continue for higher genus. At genus 3 (for $c=24$), we are at the edge of what is possible with our desktop computers. Probably to address this problem at higher genus, a new computational algorithm is necessary. The positivity conditions are linear, and the resulting spaces of solutions are convex polytopes in the space of $a_{i,j,k,..}$'s. This suggests that some sort of linear programming approach could drastically improve the efficiency of our rather basic method by searching for the edges of the space rather than enumerating every solution inside.

\begin{table}\label{c=24 table of allowed theories}
\centering
\begin{tabular}{| c | c | c | c | c | }
\hline
 $k$ & $g = 2$ & $g = 3$ & lattice & code \\ 
 \hline \hline
 1 & & & $u(1)^{24}$ &  \\
 \hline
 3 & \checkmark & \checkmark & $(a1)^{24}$ & \checkmark \\
 \hline
 4 & & & $(a2)^{12}$ & \\
 \hline
 5 & & & $(a3)^{8}$ & \\
 \hline
 6 & & &  $(a4)^6$ & \\
 \hline
 7 & \checkmark & \checkmark & $(d4)^{6}$, $(a5)^4 d4$ & \checkmark \\
 \hline
 8 & & & $(a6)^{4}$ & \\
 \hline
 9 & \checkmark & & $(a7)^{2} (d5)^{2}$ & \\
 \hline
 10 & & & $(a8)^{3}$ & \\
 \hline
 11 & \checkmark & \checkmark & $(d6)^{4}$, $(a9)^2 d6$ & \checkmark \\
 \hline
 13 & \checkmark & \checkmark & $(e6)^{4}$, $a11 \,  d7 \,  e6$ & \\
 \hline
 14 & & & $(a12)^2$ & \\
 \hline
 15 & \checkmark & \checkmark & $(d8)^{3}$ & \checkmark \\
 \hline
 17 & \checkmark & \checkmark & $a15 \, d9$ & \\
 \hline
 19 & \checkmark & \checkmark & $a17 e7$, $d10 (e7)^2$ & \checkmark \\
 \hline
 21 & \checkmark & \checkmark & & \\
 \hline
 23 & \checkmark & \checkmark & $(d12)^{2}$ & \checkmark \\
 \hline
 25 & \checkmark & \checkmark & & \\
 \hline
 26 &  &  & $a24$ & \\
 \hline
 27 & \checkmark & \checkmark & & \\
 \hline
 29 & \checkmark & \checkmark & & \\
 \hline
 31 & \checkmark & \checkmark & $d16 \, e8$, $(e8)^3$  &  \checkmark$\cdot 2$ \\
  \hline
 33 & \checkmark & \checkmark & & \\
 \hline
 35 & \checkmark & \checkmark & & \\
 \hline
 37 & \checkmark & \checkmark & & \\
 \hline
 39 & \checkmark & \checkmark & & \\
 \hline
 41 & \checkmark & \checkmark & & \\
 \hline
 43 & \checkmark & \checkmark & & \\
 \hline
 45 & \checkmark & \checkmark & & \\
 \hline
 47 & \checkmark & \checkmark & $d24$ & \checkmark \\
 \hline
 49 & \checkmark & & & \\
 \hline
 51 & \checkmark & & & \\
 \hline
 53 & \checkmark & & & \\
 \hline
 55 & \checkmark & & & \\
 \hline
 57 & \checkmark & & & \\
 \hline
 59 & \checkmark & & & \\
 \hline
 61 & \checkmark & & & \\
 \hline
 \hline
\end{tabular}
\caption{Polynomials organized by number of spin 1 currents $N_{\text{currents}}=24k$. The number of currents may be computed from the genus 1 polynomial using $N_{\text{currents}} = 72 + 16 a_4$. See \cite{Gaberdiel:2009rd} for the list of lattices organized by the number of currents.}
    \label{tab:c=24}
\end{table}

\subsubsection{Finding lattice theta functions}
\label{subsubsec:findinglatticethetas}

Above we have seen that our method may be used to find enumerator polynomials for every code that exists at $c = 24$. These define lattices, so our method is able to find some lattice theta functions. However there are other theta functions, deriving from non-code lattices, which are not captured by our approach. In general it occurs for lattices whose theta functions correspond to enumerator polynomials with half-integer coefficients and, in the case of $u(1)^{24}$, negative coefficients. We would like to understand how these lattice theta functions fit into our discussion so far.

By examining explicitly some of the theta functions for non-code lattices, we find that they can be written in enumerator polynomial form, but have fractional and negative coefficients. The specifics are quite interesting. For genus 1, there is a single lattice with negative coefficients. This is the $u(1)^{24}$ lattice, corresponding to $k = 1$ in our table. At genus 2, we find that the $k = 4$, $5$, and $6$ lattices also yield enumerator polynomials with negative coefficients. The lattices are visible as the black stars \ref{fig:allowed g2}, (b), where it is easy to see that these three polynomials are ruled out.  So it appears that even if we repeat our procedure and allow for fractional coefficients, this will still not be enough to enumerate all lattices due to this negativity.

It seems likely that the coefficients are not allowed to be arbitrarily negative, because they still need to lead to a positive state expansion. For example, for the $u(1)^{24}$ theory, has the following theta-function in polynomial form 
\begin{align}
    \Theta_{u(1)^{24}}(x_0,x_1) = x_0^{24} - 3 x_0^{20} x_1^4 + 771 x_0^{16} x_1^{8} + 2558x_0^{12}x_1^{12} + 771 x_0^8 x_1^{16}- 3 x_0^4 x_1^{20} + x_1^{24}\,.
\end{align}

We see that $a_4 = -3$, corresponding to $24$ currents. This is the smallest value of all lattices; every other lattice has $a_4>0$. It seems that what happens is that the enumerator polynomial, which represents the numerator of the partition function, is able to be slightly negative because as long as it can be compensated by the denominator (\emph{e.g.} $\eta(q)^{c}$ in the genus 1 case) to ensure that the expansion in $q$, which counts the actual state degeneracies, is purely positive. Thus it seems that arbitrary amounts of negativity are probably not allowed in the enumerator polynomial. 

It is known that there are 24 even self-dual Euclidean lattices in dimension 24 \cite{Niemeier1973}. These are a subset of the 71 meromorphic chiral CFTs \cite{Schellekens:1992db}.\footnote{The famous cases of the Leech lattice and Monster module correspond to $N_{\mathrm{currents}}=24$ and $N_{\mathrm{currents}}=0$ respectively. The Golay code gives the lattice $(a1)^{24}$ through Construction A, however through the \emph{twisted} Construction A (see \emph{e.g.}\ \cite{Dymarsky:2020qom}) it gives rise to the Leech lattice.} It would be very interesting to find the correct generalization of our procedure to enumerate lattices or general meromorphic theories, rather than just codes. 

\subsection{Summary of results}

\subsubsection{Counting polynomials with positive integer coefficients}

Having performed this procedure, we may now list our results for a variety of different values of $c$.
\begin{itemize}
    \item $c = 8$: \begin{addmargin}[2em]{0em} \begin{itemize}
        \item there is a single polynomial at genus 1, 2, and 3,
        \item these all correspond to the Hamming [8,4,4] code.
    \end{itemize}
    
    \end{addmargin}
    \item $c = 16$: \begin{addmargin}[2em]{0em} 
    \begin{itemize}
        \item at genus 1 and 2, there is a single polynomial,
        \item at genus 3, we find 1,681 polynomials, which all factorize into the unique genus 1 polynomial. These 1,681 polynomials are a linear combination of the enumerator polynomials of two different codes -- the codes associated to the $e_8 \otimes e_8$ and $d^+_{16}$ lattices -- which may be averaged in 1,681 ways to form positive integer polynomials.\footnote{As discussed in detail in section~\ref{sec:beyond}, these two codes are resolvable by the enumerator polynomials at genus 3, while in fact they have the same partition functions for all genus $g<5$. This is due to non-trivial relations between the theta constants.}
    \end{itemize}
    \end{addmargin}
    \item $c = 24$: \begin{addmargin}[2em]{0em} \begin{itemize}
        \item there are 190 genus 1 polynomials. 29 come from consistent genus 2 polynomials, and 21 at genus 3,
        \item these come from 9 codes. Note however that the genus 1 partition functions of two of the code theories agree, so that there are eight unique genus 1 code theory partition functions,
        \item we find 135,065 invariant polynomials at genus 2.
    \end{itemize} 
    \end{addmargin}
    \item $c = 32$: \begin{addmargin}[2em]{0em} \begin{itemize}
        \item we find 284 genus 1 polynomials,
        \item at genus 2, there are 210,116 polynomials with the correct factorization limits, but they only result in 161 unique genus 1 polynomials,
        \item There are 85 codes \cite{Conway1980, Conway1992}, but only 37 unique genus 1 enumerator polynomials. 80 have Hamming distance $d= 4$, and 5 have $d= 8$.
    \end{itemize}  \end{addmargin}
    \item $c = 40$: \begin{addmargin}[2em]{0em}
    \begin{itemize}
        \item there are 400 genus 1 polynomials,
        \item 246 of these come from genus 2 polynomials,
        \item we are not able to compute the total number of genus 2 polynomials
        \item there are 94,343 codes, 77,873 with Hamming distance $d=4$, the rest with $d=8$ \cite{Betsumiya2012}.
    \end{itemize}  \end{addmargin}
    \item $c = 48$: \begin{addmargin}[2em]{0em}
    \begin{itemize}
        \item there are 14,381,890 genus 1 polynomials,
        \item 2,580,972 come from genus 2 polynomials,
        \item we are not able to enumerate all genus 2 polynomials.
    \end{itemize} \end{addmargin}
\end{itemize}

\subsection{Siegel modular forms in the code variables}
We have seen that the lattice theta series $\Theta_\Lambda ( \Omega)$ transforms as modular form of weight $\frac{c}{2}$
\begin{equation}
    \Theta_\Lambda (\tilde \Omega) = \text{det} (C \Omega+D )^{\frac{c}{2}} \Theta_\Lambda\left(\Omega\right)  \, .
\end{equation}
Modular forms of even weight are generated by a ring \eqref{eq:generatorsgenus1} which allows us fix $\Theta_\Lambda $ in terms of the generators of the ring. This is possible for genus 2 and 3 as well since the generators of the ring of Siegel modular forms is known \eqref{eq:generatorsgenus2}. In this section, we constrain the space of code genus 2 CFT partitions functions and reproduce the results established in the previous subsection in this manner. The motivation for doing so is that the constraints of factorization on $\Theta_\Lambda $ can be imposed by exploiting the known factorization properties of the Siegel modular forms \eqref{Seigel forms factorization genus 2}. This was precisely the strategy used in \cite{Keller:2017iql} to constrain the space of genus 2 chiral CFTs. In this section we show that by expressing genus 2 Siegel modular forms in terms of code variables, factorization constraints can be easily put on code CFT's. 
\subsubsection{Siegel forms in code basis}
In order write the Siegel forms in terms of the code variables, one only needs to know the number of generators in the ring and the action of the Siegel $\Phi$ operator \eqref{eq:Siegel-forms-normalization}. We first consider how this works at genus 1. For $c=8$, the only allowed modular form of weight 4 is $G_4$, which fixes the associated $\Theta_\Lambda$. The associated Construction A lattice arises from the unique code of dimension 8 -- the Hamming $[8,4,4]$ code. Applying the theta map to this code gives us an expression for $G_4$ in terms of code variables.
\begin{align}\label{G4 in code vars}
   G_4 &\cong x_0^8 + 14 x_0^4 x_1^4 + x_1^8 .
\end{align}   
Similarly, at $c=24$, the only the most general  $\Theta_\Lambda$ of weight 12 is characterized by the number of spin $1$ currents $N_1$
\begin{equation}\label{c24 genus 1}
\Theta^{g=1}_\Lambda = G_4^3 +  \left( N_1-744 \right) \Delta_{12}.
\end{equation}
This statement in terms of enumerator polynomials is a well known result in coding theory and is called Gleasons theorem for binaray self dual codes. It states that the enumerator polynomial for any self dual code can be written as a linear combination of the expressions for $G_4$ and $\Delta $ written in the code basis. 
We can work backwards and express $\Delta_{12}$ in code variables since it must be given by a linear combination of $G_4^3$ in code variables and the enumerator polynomial of any even self dual $c=24$ code of choice. We then fix the overall normalization of $\Delta_{12}$ by demanding that it maps to zero under the Siegel theta map $\Phi$. The identity $\Delta_{12}=\frac{1}{1728}\left( G_4^3-G_6^2 \right)$ fixes $G_6$ in terms of code variables as well
\begin{align}
   G_6 &\cong x_0^{12} - 33 x_0^8 x_1^4 - 33 x_0^4 x_1^8 + x_1^{12} \, , \\
    \Delta_{12} & \cong  \frac{1}{16}x_0^4 x_1^4 (x_0^4 - x_1^4)^4 \, .
\end{align}
We proceed in a completely analogous fashion at genus 2.
In this case again, at $c=8$ the unique weight 4 Siegel form, $E_4$, must be the biweight enumerator polynomial of the Hamming $[8,4,4]$ code. At $c=24$, there is more freedom and the most general lattice theta series that can arise is:
\begin{equation}\label{genus 2 unfactorized c 24}
\Theta^{g=2}_\Lambda = E_4^3 +  a_1 \psi_{12} + a_2 \chi_{12}.
\end{equation}
We can take any 2 distinct $\Theta_\Lambda^{g=2}$ to be ones arising from the biweight enumerator polynomials of any two distinct $c=24$ codes. This gives us 2 equations of the form \eqref{genus 2 unfactorized c 24} which we can use to solve for $\psi_{12}$ and $\chi_{12}$.  Demanding the normalization \eqref{eq:Siegel-forms-normalization} allows us to fix the coefficients $a_1$ and $a_2$ and the end result is that can express them in code variables \eqref{eq:E4app}--\eqref{eq:chi12app}. 
The remaining Siegel form $\chi_{10}$ first appears at $c=32$ and it is fixed by repeating this procedure by taking the biweight enumerator polynomial of any  self dual $c=32$ code and expressing it terms of known Siegel forms in the code basis according to \begin{equation}\label{ genus 2 unfactorized c 32}
\Theta^{g=2}_\Lambda = E_4^4 +  a_1 E_4 \psi_{12} + a_2 E_4 \chi_{12} + a_3 E_6 \chi_{10}.
\end{equation}
The unknown constants $a_1,a_2,a_3$ are fixed by demanding that $\chi_{10} \mapsto 0 $ under the action of the Siegel $\Phi$ operator.
\subsubsection{Factorization constraints}
We have seen that under the factorization limit the period matrix becomes block diagonal and the lattice theta function factorizes
\begin{equation}
    \Theta_\Lambda^{g=2}(\Omega)\to \Theta_\Lambda^{g=2}(\tau_{1} \oplus \tau_{2})=\Theta_\Lambda^{g=1}(\tau_{1})\Theta_\Lambda^{g=1}(\tau_{2}).
\end{equation}
Using the expressions for code variables \eqref{eq:E4app}--\eqref{eq:chi12app}, we can write down putative enumerator polynomials or equivalently code CFT partition functions. For these to come from codes, we now demand that these polynomials have positive and integer degeneracy of codewords and that they factorize into squares of well defined lower genus enumerator polynomials in the factorization limit. 
\paragraph{$\boldsymbol{c=24}$}
\eqref{genus 2 unfactorized c 24} must devolve to the square of the genus 1 lattice theta function given by \eqref{c24 genus 1}. This is straightforward to implement since factorization properties of the Siegel forms are known \eqref{Seigel forms factorization genus 2}.  The result is 
\begin{equation}\label{genus 2 factorized c 24}
\Theta^{g=2}_\Lambda = E_4^3 +  \left( N_1-744 \right) \psi_{12} + \left( (N_1 - 744 ) (N_1+984) \right) \chi_{12}.
\end{equation}
It is important to emphasise that the above constraint is true in any meromorphic $c=24$ CFT and these constraints for various $c$ are discussed in \cite{Keller:2017iql}. Here, we use it to constrain the space of allowed codes factorize because we know how to express the Siegal modular forms in terms of code variables. In fact, demanding that \eqref{genus 2 factorized c 24} has positive integer coefficients when written in code variables lets us recover the result that there are only 29 enumerator polynomials which factorize at genus 2 out of the 190 consistent genus 1 code polynomials. 

\paragraph{$\boldsymbol{c=32}$}
The allowed enumerator polynomial is parametrized by the number of spin $1$ currents $N_1$:
\begin{align}
\Theta^{g=1}_\Lambda &= G_4^4 +  \left( N_1-992 \right) \Delta G_4.
\end{align}
Demanding factorization gives us 
\begin{align}
\Theta^{g=2}_\Lambda &= E_4^4 +  \left( N_1-992 \right) \Delta E_4 \psi_{12} + \left( N_1-992 \right)\left( N_1+736 \right) E_4 \chi_{12} + c_1 E_6 \chi_{10}.
\end{align}
Here $c_1$ is an unknown constant which cannot be determined by factorization constraints since $\chi_{10}\to 0$ under factorization.  For $\Theta^{g=1}_\Lambda$ to possibly arise from a code, we must have $N_1=16k$ for $6\leqslant k\leqslant 289$ and $k \in \mathbb{Z}$. In order for $\Theta^{g=2}_\Lambda$ to possibly arise from a code, the range of allowed reduces to $6\leqslant k \\leqslant  166 $. We should note however that in general each of these $k$ have multiple consistent $c_1$ associated with them obeying inequalities. The different $c_1$ correspond to allowed values of the 3 point structure coefficient $c_{i,j,k}$ of light primary operators. Imposing unitarity  does not rule out any code CFT's.
\paragraph{$\boldsymbol{c=40}$}
The allowed enumerator polynomial is parametrized by $N_1$,
\begin{align}
\Theta^{g=1}_\Lambda &= G_4^5 +  \left( N_1-1240 \right) \Delta_{12} G_4^2.
\end{align}
For $\Theta^{g=1}_\Lambda$ to possibly arise from a code, we must have $N_1=120+16k$ for $0\leqslant k\leqslant 399$ and $k \in \mathbb{Z}$. Demanding a factorizable genus 2 enumerator polynomial gives:
\begin{align}
\Theta^{g=2}_\Lambda &= E_4^5 +  (N_1 - 1240 ) E_4^2 \psi_{12} + (N_1 - 1240 ) (N_1 + 488 ) E_4^2 \chi_{12}  \\ \nonumber
&\quad+c_1 E_4 E_6 \chi_{10} + c_2 \chi_{10}^2.
\end{align}
where $c_1$ and $c_2$ are unknown coefficients.
In order for $\Theta^{g=2}_\Lambda$ to possibly arise from a code, the range of allowed reduces to $0\leqslant k \leqslant  246 $ and these are not further constrained by demanding unitarity.  
\paragraph{$\boldsymbol{c=48}$}The allowed genus 1 partition function is
\begin{align}
\Theta^{g=1}_\Lambda &=G_4^6 + \left(N_1 - 1488\right) G_4^3 \Delta_{12} + \left( N_2 - 743 N_1 +   159769 \right) \Delta_{12}^2
\end{align}
For this arise from a code, we must have $N_1=144 + 16 k_1$ and $N_2=10199 + 2160 k_1 + 256 k_2$
with $k_1$ and $k_2$ satisfying
\begin{align}
    0 \leqslant k_1 \leqslant  374 \quad\,  0 \leqslant k_2 \leqslant \frac{1}{8} \left(17296 + 733 k_1 \right)
\end{align}
or
\begin{align}
    375 \leqslant k_1 \leqslant 766 \quad\, 0 \leqslant k_2 \leqslant \frac{1}{28} \left(1997688 - 2607 k_1 \right)
\end{align}
This gives 14,381,890 genus 1 polynomials. 
\begin{align}
\Theta^{g=2}_\Lambda  &= E_4^6+E_4^3 \left(N_1 \left(N_1+238\right)-2 \left(N_2+338329\right)\right) \chi_{12} \\ \nonumber 
&\quad+E_4^3 \left(N_1-1488\right) \psi_{12}-\left(N_1+1968\right) \left(743 N_1-N_2-159769\right) \chi_{12} \psi_{12} \\ \nonumber 
&\quad+\left(-743 N_1+N_2+159769\right) \left(985 N_1+N_2+574489\right) \chi_{12}^2   \\ \nonumber
&\quad+\left(-743 N_1+N_2+159769\right) \psi_{12}^2
+ c_1 E_4^2 E_6 \chi_{10}+c_2 E_4 \chi_{10}^2.
\end{align}
Demanding that the genus 1 polynomials arise from genus 2 code polynomials gives us 2,580,972 polynomials.
\paragraph{Genus 3}
As reviewed in sections~\ref{subsec:Siegel} and \ref{sec:beyond}, starting from genus 3, there are invariant polynomials that map to zero under $\Th$. This is due to non-trivial relations among the higher genus theta functions. At genus 3, the kernel of $\Th$ is generated by the degree $16$ polynomial $j_8$. This means that we can write down a unique pre-image of the rank $4$ form $\alpha_4$
\begin{align}
    \alpha_4\cong W_{e_8}^{(3)}&= \sum_{i=0}^7x_i^8+14\sum_{i<j}x_i^4x_j^4+1344\prod_{i=0}^7
    +168\!\!\sum_{\substack{i<j<k<l\\ i+j+k+l=6,\,14,\,22 }}\!\!x_i^2x_j^2x_k^2x_l^2
     \nonumber\\
    &\quad+168\left(
    x_0^2x_1^2x_4^2x_5^2+x_0^2x_2^2x_4^2x_6^2+x_1^2x_3^2x_5^2x_7^2+x_2^2x_3^2x_6^2x_7^2
    \right),
\end{align}
but at $c\geqslant 16$ the modular forms in the code variables will only be defined up to adding terms proportional to $j_8$. For instance, the 1,681 polynomials found at $c=16$ are all of the form
\begin{equation}
\label{eq:j83implicit}
    P=\left(W_{e_8}^{(3)}\right)^2+\frac{a}{1344}j_8^{(3)}
\end{equation}
for integer $a$ in the range $-1344\leqslant a\leqslant 366$. The code $d_{16}^+$ corresponds to $a=-1344$. An explicit expression for $j_8$ is given in \eqref{eq:j8app}.

\section{Discussion}
In this paper, we have shown how code CFTs provide a simple setup to explore higher genus modular invariance: the weight-$g$ enumerator polynomial evaluated at theta constants is the numerator of the genus-$g$ partition function of a code CFT. The higher genus modular transformations take the form of linear transformations on the enumerator polynomial. It is possible to solve the constraints completely, yielding the entire set of possible code enumerator polynomials. These are further reduced by requiring 
that the partition function factorizes as the genus-$g$ Riemann surface factorizes into two lower-genus Riemann surfaces. The result is that the higher genus constraints are much more restrictive than the lower genus constraints. In the case of $c = 24$, we find 190 candidate polynomials at genus 1, 29 candidates at genus 2, and 21 candidates at genus 3 (see table ~\ref{tab:c=24}). There are exactly 9 doubly-even self-dual codes at $c = 24$ so we speculate that this process, if pushed to even higher genus, will eventually converge to 9. 

So what to make of the ruled out theories, for instance the ``fake'' genus 1 partition functions which have no corresponding partition function at genus 2 or 3? Indeed, our work was largely motivated by this question, posed in \cite{Dymarsky:2020qom} for non-chiral CFTs but with a direct counterpart in the chiral case. If the fake partition functions indeed do not correspond to any CFT, their existence poses a clear limitation of the genus 1 modular bootstrap. Here we have shown that extending the modular bootstrap to higher genus gives a significant improvement.

It is possible, however, is that the ``fake'' genus 1 partition functions may correspond to CFTs which are not derived from error-correcting codes. 
Therefore it would be desirable to enlarge the class of theories that can be captured by our approach. In section~\ref{sec:enumeration}, we investigated what relaxations are needed to find lattice theories, which may or may not come from codes. These theories have partition functions which can be written in ``enumerator polynomial form'', meaning that they are homogeneous polynomials of theta-constants. However, unlike code theories, the polynomials have coefficients which may be fractional and slightly negative. The negativity, in particular, makes it difficult to devise a finite algorithm to enumerate them. However, they do not seem to be allowed to be arbitrarily negative, which leaves open the possibility that it may be possible to enumerate them by only slightly relaxing the positivity constraint. It is natural to try this at $c=24$, where there are 24 even self-dual Euclidean lattices \cite{Niemeier1973}, and 71 meromorphic theories in total \cite{Schellekens:1992db}.

Another interesting direction for future work would be to try to make contact with the traditional bootstrap program \cite{Rattazzi:2008pe}. Higher genus partition functions contain some information about OPE coefficients, so some constraints on dynamical information are implied by higher genus modular invariance. It would be interesting to try to understand the extent of this -- in particular, are the constraints implied by modular invariance at all genera equivalent to those implied by consistency of sphere $n$-point functions with crossing, or is one constraint stronger than the other? It would also be interesting to see in detail how the factorization limits, which go beyond the symmetries of the theory, might be related to OPE data. Specifically, do the higher genus partition functions which do not factorize correctly lead to OPE data which is somehow pathological? In this paper, we considered the complete factorization limit, where $ \Omega$ is made strictly block diagonal. However one can consider subleading corrections to this factorization, taking off-diagonal elements of $\Omega$ to be small expansion parameters. This yields information about averages of OPE coefficients for light operators \cite{Keller:2017iql}. We leave exploring this systematically to future work. 

Our work opens avenues to explore the relationship between quantum error-correcting codes, Lorentzian lattices, and (non-chiral) CFTs \cite{Dymarsky:2020bps, Dymarsky:2020qom,Dymarsky:2021xfc} at higher genus. As we will discuss in \cite{Future}, this relationship admits a higher genus generalization similar to the one presented in this paper. The enumerator polynomial of a quantum error-correcting code at genus $g$ is a polynomial in $2^{g-1}(2^g+1)=3,10,36,\ldots$ variables. This polynomial evaluated at theta constants gives the higher genus partition function of a putative non-chiral CFT. The constraints on higher genus partition functions from modular invariance and factorization are interesting and largely unexplored areas, and the techniques developed in this paper give clear directions to understand them. This will be explored in a future paper \cite{Future}.

Another interesting byproduct of our work is a set of explicit expressions for genus 2 and 3 Siegel modular forms in terms of the polynomials of theta constants. These expressions facilitate easy manipulation of the forms -- for example, it is obvious in our basis if a form is a product of two other forms. In principle, one could use our method to determine the full ring of modular forms for genus $g \leqslant 3$. We believe that this approach would be particularly interesting in the case of quantum error-correcting codes, which correspond to non-chiral CFTs and where the theory of modular forms is much less developed. 

Interesting recent work has given a possible holographic bulk interpretation to the average over Narain lattice CFT's \cite{Maloney:2020nni, Afkhami-Jeddi:2020ezh}, which are computed using the Siegel-Weil formula. The chiral version of this formula averages even unimodular Euclidean lattices of dimension $c$, to give a (holomorphic) Siegel modular form at every genus, \emph{e.g.} see \cite{Zagier2008},
\begin{align}
    \sum_{\Lambda}\frac{1}{\mathrm{Aut}(\Lambda)}\Theta_{\Lambda}= m_{\frac{c}{2}} E^{(g)}_{\frac{c}{2}} \, ,
\end{align}
where $E_k^{(g)}$ is the genus $g$ holomorphic Eisenstein series.
Here the sum is weighted by the number of automorphisms of each lattice and  $m_k=\frac{B_k}{2k}\frac{B_2}{4}\frac{B_4}{8}\cdots\frac{B_{2k-2}}{4k-4}$ where $B_k$ denotes the $k^{th}$ Bernoulli number. 
 The appearance of the Siegel modular form suggests that a bulk interpretation of the chiral average may also be possible \cite{Dymarsky:2020pzc}. In this context, it would be interesting to see if our discussion on factorization of individual lattice partition functions provides a simple setting to study statistical properties of the averaged partition function.

\section*{Acknowledgments} We thank Anatoly Dymarsky for very interesting early conversations which ultimately led to this project. We also thank R. Volpato and R. Salvati Manni for useful clarifications and A. Vichi for valuable comments in the draft. JH thanks E. Hamilton and M. Elmi for useful discussions. The work by JH and BM has received funding from the European Research Council (ERC) under the European Union's Horizon 2020 research and innovation program (grant agreement no.~758903). AK received partial support from the National Science Foundation under Grant No. NSF PHY-1748958.

\appendix
\label{sec:quantum codes}

\section{Conventional Genus 2 Transformations}
For convenience, we provide the more conventional set of $\mathrm{Sp}(4, \mathbb{Z})$ transformations in terms of their action on our code variables. These are given by six elements which act on the period matrix according to
\begin{align}
    &T_1 : \Omega_{11}\ \mapsto\  \Omega_{11}+1, \quad \quad 
    T_2 : \Omega_{22}\ \mapsto\  \Omega_{22}+1 ,\label{genus2 T}
\end{align}

\begin{align}
    &S_1: \begin{cases}
        \Omega_{11} & \mapsto \quad -1/\Omega_{11} ,\\
        \Omega_{12} & \mapsto \quad -\Omega_{12} / \Omega_{11}, \\
        \Omega_{22} & \mapsto \quad \Omega_{22} -\Omega_{12}^2 / \Omega_{11} \nonumber
    \end{cases}
    &S_2: \begin{cases}
        \Omega_{22} & \mapsto \quad -1/\Omega_{22} ,\\
        \Omega_{12} & \mapsto \quad -\Omega_{12} / \Omega_{22} ,\\
        \Omega_{11} & \mapsto \quad \Omega_{11} -\Omega_{12}^2 / \Omega_{22} ,\label{genus2 S}
    \end{cases} 
\end{align}
\begin{align}
    &V: \Omega_{12}\ \mapsto\ -\Omega_{12} , \\  
    &U: \Omega_{12}\ \mapsto\ \Omega_{12} + 1 .\label{genus2 U V}
\end{align}
$\langle S_1,T_1 \rangle $ and $\langle S_2,T_2 \rangle $ are the $\mathrm{Sp}(2,Z)$ subgroups which generate modular transformations for the two tori. These transformations lead to simple linear transformations on the code variables:
\begin{align}
\begin{split}
    T_1& : \quad x_0   \mapsto  i x_0, \quad 
    x_2 \mapsto  i x_2, \quad 
    x_1 \mapsto   x_1, \quad 
    x_3 \mapsto  x_3 ;\\
    T_2& : \quad x_0  \mapsto x_0, \quad 
    x_2 \mapsto x_2, \quad 
    x_1 \mapsto i x_1, \quad 
    x_3 \mapsto  i x_3; \\
    U& : \quad x_0 \mapsto - x_0, \quad 
    x_2 \mapsto x_2, \quad 
    x_1 \mapsto  x_1, \quad 
    x_3 \mapsto   x_3;
\end{split}
\end{align}
and
\begin{align}
    S_1& : \quad x_0 \mapsto \frac{x_2 - x_0}{\sqrt 2},  \quad
    x_2 \mapsto \frac{x_2 + x_0}{\sqrt 2}, \quad x_1 \mapsto \frac{x_3 - x_1}{\sqrt 2} , \quad x_3 \mapsto \frac{x_3 + x_1}{\sqrt 2} ;\\
    S_2& : \quad x_0 \mapsto \frac{x_1 - x_0}{\sqrt 2},  \quad
    x_1 \mapsto \frac{x_1 + x_0}{\sqrt 2}, \quad x_2 \mapsto \frac{x_3 - x_2}{\sqrt 2} , \quad x_3 \mapsto \frac{x_3 + x_2}{\sqrt 2} \,.
\end{align}

\section{Siegel Modular Forms}

\subsection{Forms in the code basis}
\label{app:modularformsincodebasis}

We consider Siegel modular forms for the group $\mathrm{Sp}(2g,\Z)$. First recall from section~\ref{sec:review} that we have
\begin{align}
\begin{split}
    G_4 &\cong x_0^8 + 14 x_0^4 x_1^4 + x_1^8 ,\\
   G_6 &\cong x_0^{12} - 33 x_0^8 x_1^4 - 33 x_0^4 x_1^8 + x_1^{12} \, , \\
    \Delta_{12} & \cong  \frac{1}{16}x_0^4 x_1^4 (x_0^4 - x_1^4)^4 \, .
\end{split}
\end{align}

Note also the alternative ways \eqref{eq:G4alt} of writing the forms $G_4$ and $\Delta_{12}$. In the code basis, the modular forms at genus 2 are given by
\begin{equation}
\label{eq:E4app}
    E_4 \cong x_0^8+14 x_0^4 x_2^4+14 x_0^4 x_1^4+14 x_0^4 x_3^4+168 x_0^2 x_2^2 x_1^2 x_3^2+x_2^8+14 x_2^4 x_1^4+14 x_2^4 x_3^4+x_1^8+14 x_1^4 x_3^4+x_3^8  ,
\end{equation}
\begin{align}
\begin{split}
    E_6 &\cong x_0^{12}-33 x_0^8 x_2^4-33 x_0^8 x_1^4-33 x_0^8 x_3^4+792 x_0^6 x_2^2 x_1^2 x_3^2-33 x_0^4 x_2^8+330 x_0^4 x_2^4 x_1^4 \\  
    &\quad+330 x_0^4 x_2^4 x_3^4-33 x_0^4 x_1^8+330 x_0^4 x_1^4 x_3^4-33 x_0^4 x_3^8+792 x_0^2 x_2^6 x_1^2 x_3^2 \\  
    &\quad+792 x_0^2 x_2^2 x_1^6 x_3^2+792 x_0^2 x_2^2 x_1^2
   x_3^6+x_2^{12}-33 x_2^8 x_1^4-33 x_2^8 x_3^4-33 x_2^4 x_1^8+330 x_2^4 x_1^4 x_3^4 \\  
    &\quad-33 x_2^4 x_3^8+x_1^{12}-33 x_1^8 x_3^4-33 x_1^4 x_3^8+x_3^{12},
    \label{eq:E6app}
\end{split}
\end{align}
\begin{align}
\begin{split}
    \chi_{10} \cong \frac{1}{256} \Big(&x_0^{14} x_2^2 x_1^2 x_3^2-x_0^{12} \Big(x_2^4 \Big(x_1^4+x_3^4\Big)+x_1^4 x_3^4\Big) -x_0^{10} x_2^2 x_1^2 x_3^2 \Big(x_2^4+x_1^4+x_3^4\Big)\\  
&+x_0^8 \Big(2 x_2^8
   \Big(x_1^4+x_3^4\Big)+x_2^4 \Big(2 x_1^8+13 x_1^4 x_3^4+2 x_3^8\Big) +2 x_1^4 x_3^4 \Big(x_1^4+x_3^4\Big)\Big)\\  
& -x_0^6 x_2^2 x_1^2 x_3^2 \Big(x_2^8+14 x_2^4
   \Big(x_1^4+x_3^4\Big)+x_1^8+14 x_1^4 x_3^4+x_3^8\Big) \\  
&-x_0^4 \Big(x_2^{12} \Big(x_1^4+x_3^4\Big)-x_2^8 \Big(2 x_1^8+13 x_1^4 x_3^4+2 x_3^8\Big)+x_1^4 x_3^4 \Big(x_1^4-x_3^4\Big)^2 \\ 
&\quad+x_2^4
   \Big(x_1^{12}-13 x_1^8 x_3^4-13 x_1^4 x_3^8+x_3^{12}\Big)\Big)+x_0^2 x_2^2 x_1^2 x_3^2 \Big(x_2^{12}-x_2^8
   \Big(x_1^4+x_3^4\Big) \\  
&-x_2^4 \Big(x_1^8+14 x_1^4 x_3^4+x_3^8\Big)+\Big(x_1^4-x_3^4\Big)^2 \Big(x_1^4+x_3^4\Big)\Big)\\  
&-x_2^4 x_1^4 x_3^4 \Big(x_2^8-2 x_2^4
   \Big(x_1^4+x_3^4\Big)+\Big(x_1^4-x_3^4\Big)^2\Big)\Big),
   \label{eq:chi10app}
\end{split}
\end{align}
\begin{align}
\begin{split}
    \chi_{12} &\cong \frac{1}{768} \Big (x_0^{18} x_2^2 x_1^2 x_3^2+2 x_0^{16} \Big(x_2^4 \Big(x_1^4+x_3^4\Big)+x_1^4 x_3^4\Big)-12 x_0^{14} x_2^2 x_1^2 x_3^2 \Big(x_2^4+x_1^4+x_3^4\Big)\\  
    &\quad-2 x_0^{12} \Big(x_2^8 \Big(x_1^4+x_3^4\Big)+x_2^4
   \Big(x_1^8-38 x_1^4 x_3^4+x_3^8\Big)+x_1^4 x_3^4 \Big(x_1^4+x_3^4\Big)\Big) \\  
    &\quad+2 x_0^{10} x_2^2 x_1^2 x_3^2 \Big(11 x_2^8-26 x_2^4 \Big(x_1^4+x_3^4\Big) +11 x_1^8-26 x_1^4 x_3^4+11 x_3^8\Big) \\ 
    &\quad-2 x_0^8 \Big(x_2^{12} \Big(x_1^4+x_3^4\Big)-18 x_2^8 \Big(x_1^8+x_1^4 x_3^4+x_3^8\Big) +x_2^4 \Big(x_1^{12}-18 x_1^8 x_3^4-18 x_1^4 x_3^8+x_3^{12}\Big) \\ 
    &\quad+x_1^4 x_3^4 \Big(x_1^8-18 x_1^4 x_3^4+x_3^8\Big)\Big) -4 x_0^6 x_2^2 x_1^2 x_3^2 \Big(3 x_2^{12}+13  x_2^8 \Big(x_1^4+x_3^4\Big)\\  
    &\quad+x_2^4 \Big(13 x_1^8+2 x_1^4 x_3^4+13 x_3^8\Big) +3 x_1^{12}+13 x_1^8 x_3^4+13 x_1^4 x_3^8+3 x_3^{12}\Big) \\  
    &\quad+2 x_0^4 \Big(x_2^{16} \Big(x_1^4+x_3^4\Big)-x_2^{12} \Big(x_1^8-38 x_1^4
   x_3^4+x_3^8\Big)-x_2^8 \Big(x_1^{12}-18 x_1^8 x_3^4-18 x_1^4 x_3^8+x_3^{12}\Big) \\
    &\quad+x_2^4 \Big(x_1^{16}+38 x_1^{12} x_3^4+18 x_1^8 x_3^8+38 x_1^4 x_3^{12}+x_3^{16}\Big) +x_1^4 x_3^4 \Big(x_1^4-x_3^4\Big)^2 \Big(x_1^4+x_3^4\Big)\Big) \\  
    &\quad+x_0^2 x_2^2 x_1^2 x_3^2 \Big(x_2^{16}-12 x_2^{12} \Big(x_1^4+x_3^4\Big)+x_2^8 \Big(22 x_1^8-52 x_1^4 x_3^4+22 x_3^8\Big) \\  
    &\quad-4 x_2^4 \Big(3 x_1^{12}+13 x_1^8 x_3^4+13 x_1^4 x_3^8+3 x_3^{12}\Big)+\Big(x_1^4-x_3^4\Big)^2 \Big(x_1^8-10 x_1^4 x_3^4+x_3^8\Big)\Big) \\  
    &\quad+2 x_2^4 x_1^4 x_3^4 \Big(x_2^{12}-x_2^8 \Big(x_1^4+x_3^4\Big) -x_2^4 \Big(x_1^8-18 x_1^4 x_3^4+x_3^8\Big) +\Big(x_1^4-x_3^4\Big)^2 \Big(x_1^4+x_3^4\Big)\Big)\Big).
\label{eq:chi12app}
\end{split}
\end{align}
The normalization of $\chi_{10}$ has been fixed by matching with the alternative formula
\begin{equation}
    \chi_{10}=2^{-14}\prod_{\mathbf A}\theta[\mathbf A](\Omega)^2,
\end{equation}
where the product is over the 10 vectors of the form $(a_1,a_2,b_1,b_2)^{T}$, $a_i,b_i\in\{0,1/2\}$ that give a non-zero theta constant, see comments after \eqref{eq:thetaford8kplus}.

At genus $3$ we find (in the notation of \cite{Runge1996})
\begin{align}
    \alpha_4&\cong (8)+14(4,4)+168(2,2,2,2)+1344(1,1,1,1,1,1,1,1),
    \\
    j_8&=1344\big[2(9,1,1,1,1,1,1,1)+(8,0,0,0,2,2,2,2)+(6,2,2,2,4,0,0,0)\nonumber\\&\quad\qquad\ \, +4(5,5,1,1,1,1,1,1)-(4,4,4,0,4,0,0,0)-2(4,4,0,0,2,2,2,2) \label{eq:j8app}\\&\quad\qquad\ \, +16(3,3,3,3,1,1,1,1)-72(2,2,2,2,2,2,2,2)\big].\nonumber 
\end{align}

\subsection{Holomorphic Eisenstein series and modular forms}
\label{eq:holoeisenstein}

Here we will provide a few details on the holomorphic Eisenstein series which might be useful for computating the modular forms. At general genus, can be defined as
\begin{equation}
    E_k^{(g)}=\sum_{C,D} \big(\det(C\Omega+D)\big)^{-k}, \qquad \begin{pmatrix}
   A & B \\ C & D  
    \end{pmatrix}\in \mathrm{Sp}(2g,\mathbb Z),
\end{equation}
however there are different normalization conventions used in the literature.

For genus 1 and 2, the holomorphic Eisenstein series generates the entire ring of modular forms \cite{Igusa1964}. For genus $g\geqslant3$ this is no longer the case, since there are modular forms which cannot be written as a polynomial of the Eisenstein series \cite{Tsuyumine1986}. For the genus $g\leqslant3$, the ring of modular forms has the following generators,
\begin{align}
    & g=1 & & G_4,\ G_6,
    \label{eq:formsgenus1}
    \\
    & g=2 && E_4,\ E_6,\ \chi_{10},\ \chi_{12},
    \label{eq:formsgenus2}
    \\
    \label{eq:formsgenus3}
    & g=3 && \begin{matrix*}[l]\alpha_4, \ \alpha_6,\ \alpha_{10},\, \alpha_{12},\, \alpha'_{12},\, \beta_{14},\, \alpha_{16},\, \beta_{16},\, \chi_{18},\, \alpha_{18},\, \\\alpha_{20},\, \gamma_{20},\, \beta_{22},\, \beta'_{22},\, \alpha_{24},\,\gamma_{24},\,\gamma_{26},\, \chi_{28},\, \alpha_{30},\end{matrix*}
\end{align}
where there are $19$ generators at genus $g=3$ \cite{Lercier2019}.  \cite{Tsuyumine1986} originally gave $34$, however, \cite{Lercier2019} show that there are $19$ generators among these $34$ forms forms.
The case at genus $g=2$ was demonstrated in \cite{Huffman1979}, and related to codes in \cite{Runge1993} (using references in \cite{Runge1996}). The general theorem, that the (pseudo)reflections on codes generate the ring of modular form was done in \cite{Runge1996}. 
The set of generators at genus $4$ is not known \cite{Oura2008,Lercier2019}.

\subsubsection*{Degree one}

At degree $1$ we follow \cite{Keller:2017iql} and define
\begin{equation}
    G_k(e^{2\pi i \tau})=\frac1{2\zeta(k)}\sum_{(c,d)\in\mathbb Z^2\setminus\{(0,0)\}}\frac1{(c\tau+d)^k}.
\end{equation}
This evaluates to
\begin{equation}
    G_k(q)=1+\frac2{\zeta(1-k)}\sum_{n=1}^\infty \sigma_{k-1}(n)q^n,
\end{equation}
where $\sigma_k(n)$ is the divisor sum function defined by
\begin{equation}
    \sigma_k(n)=\sum_{d|n}d^k.
\end{equation}
The basis is generated by two elements, $G_4$ and $G_6$. In terms of these,
\begin{align}
G_8&=G_4^2\,,
\\
G_{10}&=G_4G_6\,,
\\
G_{12}&=\frac{441}{691}G_4^3+\frac{250}{691}G_6^2\,,
\\
G_{14}&=G_4^2G_6\,,
\\
G_{16}&=\frac{1617}{3617}G_4^4+\frac{2000}{3617}G_4G_6^2\,,
\\
G_{18}&=\frac{38367}{43867}G_4^3G_6+\frac{5500}{43867}G_6^3\,,
\end{align}
and 
\begin{equation}
    \Delta_{12}=\frac1{1728}(G_4^3-G_6^2).
\end{equation}
Notice that there is a direct relation 
\begin{equation}
    \Delta_{12}(e^{2\pi i\tau})=\eta(\tau)^{24}\,,
\end{equation}
where $\eta(\tau)$ is the Dedekind eta function defined in \eqref{eq:dedekindeta}.

\subsubsection*{Degree two}

The ring of degree two modular forms are generated by four elements, $E_4$, $E_6$, $\chi_{10}$, and $\chi_{12}$, which are given in terms of code variables in equations \eqref{eq:E4app}-\eqref{eq:chi12app}.  We may write the holomorphic Eisenstein series in terms of these:
\begin{align}
    E_8 &=E_4^2 \,,    \label{eq: Eki}
    \\
    E_{10} &=E_4E_6-\frac{9231667200}{43867}\chi_{10}\,,
    \\
    E_{12} &=\frac{441}{691}E_4^3+\frac{250}{691} E_6^2-\frac{36980665344000}{53678953} \chi _{12}\,,
    \\
    E_{14} &=E_4^2 E_6-\frac{187420262400 }{657931}E_4\chi_{10}\,,
    \\\nonumber
    E_{16} &=\frac{1617}{3617}E_4^4+\frac{2000 }{3617}E_4E_6^2-\frac{4600443838734729216000}{6232699579062017}E_4\chi_{12}\\&\quad-\frac{473779992577941504000}{6232699579062017}E_6\chi_{10}\,,
    \\\nonumber
    E_{18}&=\frac{38367}{43867}E_4^3E_6+\frac{5500} {43867}E_6^3-\frac{1688190624014720716800}{6651496075469717}E_4^2\chi_{10}\\&\quad -\frac{2177976079791654912000 }{6651496075469717}E_6\chi_{12}\,.
    \label{eq: Ekf}
\end{align}

We shall briefly describe the implementation of the holomorphic Eisenstein series, which was also reviewed in \cite{Keller:2017iql} following \cite{vanderGeer2008}. These are given in terms of the moduli $\Omega$ by 
\begin{align}
    E_k(\Omega) = \sum_{m, n = 0}^\infty \sum_{p^2 \leqslant 4 m n} a_k(m, n, p) q^n r^p s^m\,,
\end{align}
where $q$, $r$, and $s$ are defined in \eqref{eq:qrs}, the sum over $p$ includes all integers (including negative ones) satisfying $p^2\leqslant 4mn$, and the Fourier coefficients $a_k$ are given by 
\begin{align}
    a_k(m,n,p) \ = \ \frac{2}{\zeta(3-2k) \zeta(1-k)} \sum_{d|(n,m,p)} d^{k-1} H\left(k-1, \frac{4nm-p^2}{d^2} \right)\,,
\end{align}
where the sum is over the divisors $d$ of the GCD of $n$, $m$, and $p$. $H(r, N)$ is the Cohen function, defined in \cite{Cohen1975} as
\begin{align}
    H(r, N) = \begin{cases} 
      \sum_{d^2|N} h(r, N/d^2) & (-1)^rN \equiv 0 \text{ or } 1 \text{ (mod 4) }, \\
      \zeta(1 - 2r) & N = 0, \\
      0 & \text{otherwise} \, . 
   \end{cases}
\end{align}
The sum in the first term is over all squares $d^2$ which divide $N$. The function $h(r, N)$ is then defined by 
\begin{align}
    h(r,N) = \begin{cases} 
      (-1)^{\left \lfloor{r/2}\right \rfloor } (r-1)! N^{r - 1/2} 2^{1-r} \pi^{-r} L(r, \chi_{(-1)^r N}), & (-1)^rN \equiv 0 \text{ or } 1 \text{ (mod 4) }, \\
      0, &  (-1)^rN \equiv 2 \text{ or } 3 \text{ (mod 4) } . \end{cases}
\end{align}
Finally, the character $\chi_{D}(d)$ can be written using the Kronecker symbol via
\begin{align}
    \chi_{D}(d)= \left( \frac{D}{d} \right) \, ,
\end{align}
which allows us to compute the $L-$function 
\begin{align}
    L(r, \chi_{(-1)^r N}) = \sum_{n = 1}^\infty \left( \frac{(-1)^r N}{n} \right) n^{-r} \, .
\end{align}
The result of this chain of computations is an algorithm for computing the Eisenstein series $E_k$. In practice, we could not compute the infinite sums associated with the $L$-functions exactly, so we computed them to large values of $n$ and determined the formulas \eqref{eq: Eki}--\eqref{eq: Ekf} numerically.


\bibliography{cite.bib}

\bibliographystyle{JHEP.bst}

\end{document}